**Yu.N.Efremov[1], V.L.Afanasiev[2], E.J.Alfaro,[3] R.Boomsma[4], N.Bastian[5], S. Larsen [6], M.C.Sánchez-Gil[3], O.K.Silchenko[1], B.García-Lorenzo[7], C.Muñoz-Tuñon [7], and P.W.Hodge[8]**

# Ionized and neutral gas in the peculiar star/cluster complex in NGC 6946

*1 Sternberg Astronomical Institute, MSU, Moscow 119992, Universitetsky pr. 13, RF*
*2 Special Astrophysical Observatory, RAS, Nizhny Arkhyz, 369167, RF*
*3 Instituto de Astrofísica de Andalucía, CSIC, Apartado 3004, Granada, Spain*
*4 Kapteyn Astronomical Institute, University of Groningen, P.O. Box 800, 9700 AV, Groningen, The Netherlands*
*5 Department of Physics and Astronomy, University College London, Gower Street, London, WC1E 6BT, UK*
*6 Astronomical Institute, University of Utrecht, Princetonplein 5, NL-3584 CC, Utrecht, The Netherlands*
*7 Instituto de Astrofísica de Canarias, E-38200 La Laguna, Tenerife, Spain*
*8 Department of Astronomy, University of Washington, Seattle, WA 98195, USA*

**Abstract**

The characteristics of ionized and HI gas in the peculiar star/cluster complex in NGC 6946, obtained with the 6-m telescope (BTA) SAO RAS, the Gemini North telescope, and the Westerbork Synthesis Radio Telescope (WSRT), are presented. The complex is unusual as hosting a super star cluster, the most massive known in an apparently non-interacting giant galaxy. It contains a number of smaller clusters and is bordered by a sharp C-shaped rim. We found that the complex is additionally unusual in having peculiar gas kinematics. The velocity field of the ionized gas reveals a deep oval minimum, ~300 pc in size, centered 7" east of the supercluster. The $V_r$ of the ionized gas in the dip center is 100 km/s lower than in its surroundings, and emission lines within the dip appear to be shock excited. This dip is near the center of an HI hole and a semi-ring of HII regions. The HI (and less certainly, HII) velocity fields reveal expansion, with the velocity reaching ~30 km/s at a distance about 300 pc from the center of expansion, which is near the deep minimum position. The super star cluster is at the western rim of the minimum. The sharp western rim of the whole complex is plausibly a manifestation of a regular dust arc along the complex edge. Different hypotheses about the complex and the $V_r$ depression's origins are discussed, including a HVC/dark mini-halo impact, a BCD galaxy merging, and a gas outflow due to release of energy from the supercluster stars.

## 1 INTRODUCTION

A bright isolated stellar complex shines in the outskirts of the spiral galaxy NGC 6946, about 5 kpc southwest of the galaxy centre. It was first reported by Hodge (1967), as an object somewhat similar to the system of the giant stellar arcs he had noted in the LMC. Thirty-two years later this object was independently found by Larsen & Richtler (1999) with the Nordic Optical Telescope (NOT); they described the complex as a round bright concentration of stars and clusters about 600 pc in diameter. They find one of the clusters to be outstanding by its large size and brightness. The complex is remarkable also because of its sharp arc-like western edge, which is a 130-degree long segment of a regular circle (Fig. 1).

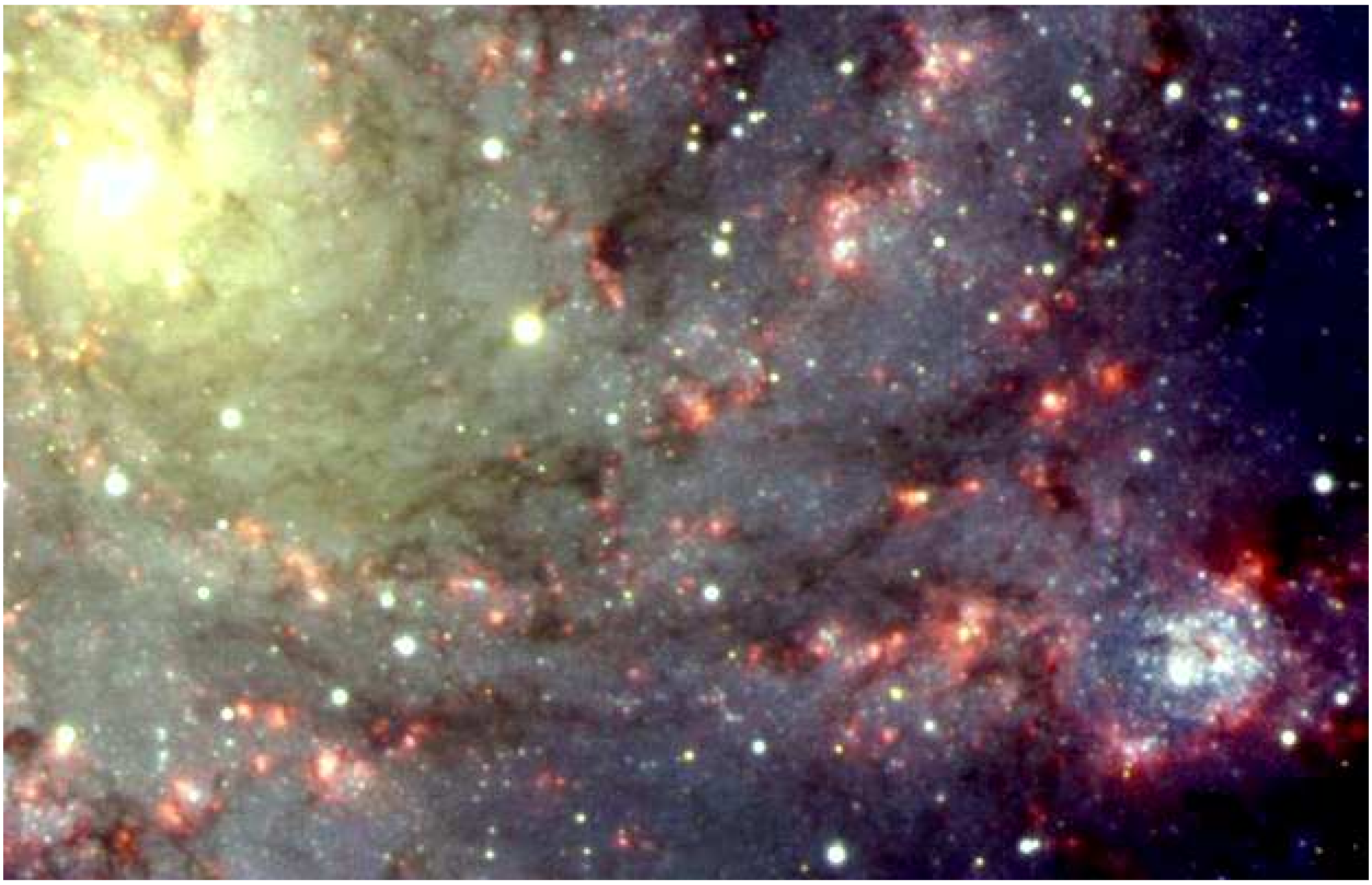

Figure 1. The peculiar complex in NGC 6946. The detail of image obtained at the Gemini North telescope (Gemini Observatories/AURA). The diameter of the complex is 22”. North at top, East to the left.

The properties of the star clusters in the complex have been first studied by Elmegreen et al. (2000) based on UBV photometry obtained at the NOT. Later on the complex was investigated in more detail with the HST and Keck-I telescopes by Larsen et al. (2001, 2002), who studied the star formation history in the complex and determined the velocity dispersion inside the giant cluster, its age (about 13 Myr) and luminosity (Mv = ~- 13.2). Its photometric and dynamical masses were also estimated from these observations; both estimations provide similar values close to $10^6$ solar masses. It was possible therefore to suggest that the cluster is gravitationally bound and that it resembles a young globular cluster, eventually to evolve into an old classical globular cluster. Its parameters fit those for giant young compact clusters, often called super star clusters (SSC).

Such clusters are commonly found in interacting galaxies and have been also detected in central areas of galaxies with active star formation (mostly post-merging) as well as in dwarf compact galaxies. However, its location in the isolated complex of peculiar shape in the outskirts of a normal spiral galaxy makes this object peculiar. Only half a dozen similar clusters are known in such galaxies, most of these being less massive than the NGC 6946 cluster (Larsen & Richtler, 2004).

The density of high luminosity stars and clusters within the Hodge complex is so high that it may be classified as a localized burst of recent star formation, the maximum rate of star formation there being some 30 Myr ago (Larsen et al., 2002). Such star/cluster complexes (the younger of which are known also as super associations) are very rare (one per cent of common complexes) in normal galaxies and some special conditions are plausibly needed to trigger a burst of star formation there (Efremov, 2004a,b). A smaller version of the Hodge complex is known in M51, it is the G1 complex which is round and rather isolated, but large clusters are missing there (Bastian et al., 2005).

Long slit spectroscopy along three selected directions was obtained with the 6-m BTA and 10-m Keck-I telescopes (Efremov et al., 2002). The observed radial velocity curves along the slits showed several dips and bumps superimposed on a smooth curve, compatible with the expected rotation curve for this region of the galaxy. The most intriguing feature of the spectra is the steep fall-off in the V(r) curve at PA 83°, with a center 7" East of the SSC, in the region of the lowest brightness of Hα. It was interpreted as the outcome of an expanding semi-bubble. This feature has added one more peculiarity to the unique properties of the complex.

The origin and evolution of the complex and its almost central SSC cannot be resolved without more detailed data on the kinematics and physics of the gas in and around the complex. To this end 3D spectroscopy of the ionized gas was carried out with the 6m BTA telescope of the Special Astrophysical Observatory of the Russian Academy of Sciences. Observations and reduction have been done by V.L.Afanasiev. Data for HI gas in the region were also obtained, done at the Westerbork Synthesis Radio Telescope (WSRT) by R.Boomsma within the framework of studies of the whole galaxy. Some results obtained by E.J.Alfaro and B.García-Lorenzo with the 4m WHT telescope (Roque de los Muchachos Observatory), N.Bastian at the Gemini North telescope, and by S.Larsen at the Keck-II telescope are also presented.

## 2 BTA observations and reductions

In August 2003 the complex was observed at the primary focus of the 6-m telescope (BTA), using the Multi-Pupil Fiber Spectrograph (MPFS) (Afanasiev et al. 2001). This spectrograph gives simultaneously spectra from 256 rectangular cells, 1" x 1" in size each, arranged as a rectangular matrix 16 x 16 squared arcsec. The detector was a CCD EEV42-40 (2048 x 2048 pixels). A detailed description of the spectrograph can be found on the SAO web-page (http://www.sao.ru/hq/lsfvo/devices/mpfs/mpfs_main.html).

A mosaic of four frames centered on the SSC position was obtained together with two more frames at east and north-east locations relative to the SSC center. The logbook of the observations is shown in Table 1, and the locations of the fields can be seen in Figs. 2 and 3.

Table 1. Journal of observations

| Date | position | Offset<br>RA DEC | Spectral<br>coverage Å | Dispersion<br>Å/px | exposure<br>time (s) | Seeing (“) |
|---|---|---|---|---|---|---|
| 26 aug 2003 | SW | -5.5” -6.0” | 3840-6810 | 1.45 | 3x1200 | 2.5 |
| 26 aug 2003 | NE | -6.5 +5.5 | 3840-6810 | 1.45 | 3x1200 | 2 |
| 26 aug 2003 | NW | +6.0 +6.5 | 3840-6810 | 1.45 | 3x1200 | 1.5 |
| 26 aug 2003 | SW | +6.0 -5.0 | 3840-6810 | 1.45 | 3x1200 | 1.5 |
| 24 aug 2003 | SW | -5.5 -6.0 | 6140-7120 | 0.76 | 3x900 | 1.2 |
| 24 aug 2003 | NE | -6.5 +5.5 | 6140-7120 | 0.76 | 3x900 | 1.2 |
| 24 aug 2003 | NW | +6.0 +6.5 | 6140-7120 | 0.76 | 3x900 | 1.2 |
| 24 aug 2003 | SW | +6.0 -5.0 | 6140-7120 | 0.76 | 3x900 | 1.2 |
| 29 aug 2003 | E1 | -22.5 +24 | 6140-7120 | 0.76 | 3x900 | 1.5 |
| 29 aug 2003 | E2 | -22.5 0 | 6140-7120 | 0.76 | 3x900 | 1.5 |

The coordinates of the centers of each field are given relative to the SSC center coordinates (RA $20^h34^m 32^s.04$; DEC +60º 08' 14".6; 2000.0). The observations were conducted with two diffraction gratings: 600 rulings/mm (dispersion 1.45 Å/px) and 1200 rulings/mm (dispersion 0.76 Å/px). The first grating was used for 2D-spectrophotometery in the studied fields, and the second for analysis of the velocity field. For flux calibration the spectrophotometric standard stars G24-9 and BD+25º 4655 were used.

The reduction of the data was conducted by a standard mode (bias subtraction, cleaning for cosmic rays, correcting of geometrical and photometric distortion, wavelength and absolute flux calibrations, sky background subtraction) through the software package MPFS_2K.lib, working in IDL. For every region, intensity line maps, intensity line ratio maps and velocity field maps were constructed from the cube of data. It is necessary to note, that because of the low line-of-sight systemic velocity of NGC 6946, apart from the typical lines of hydroxyl, the noticeable line of telluric Hα is very close to the emission coming from the target galaxy. To accurately remove this annoying sky line we have used spectra from a separate field, distant far enough from the galaxy (30'). Once a single spectrum of the sky was obtained, it was subtracted from the 2D-spectra of the targets. The profile of the Hα emission line within the complex was thus correctly isolated and defined.

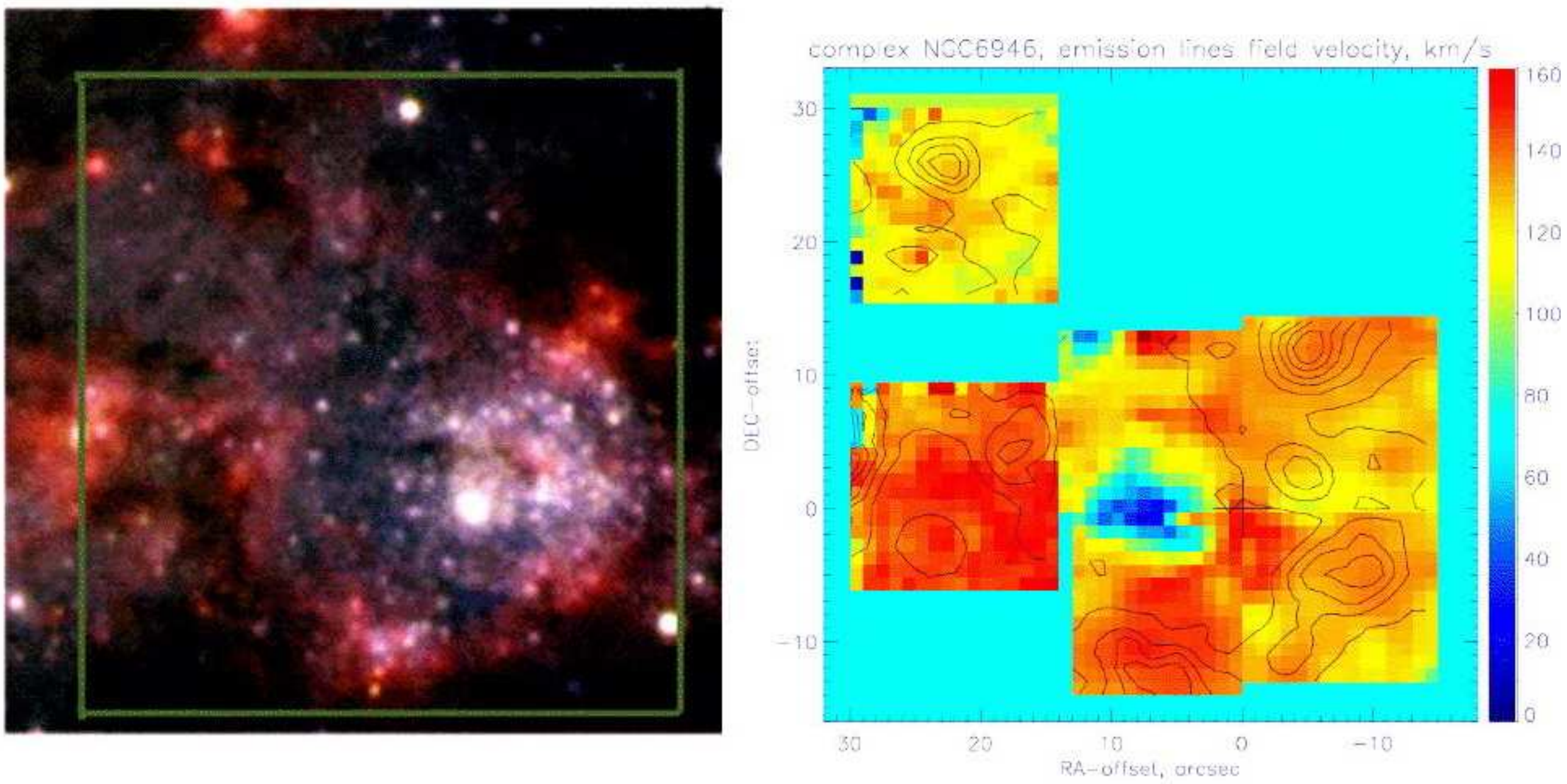

Figure 2. Left: the Gemini North image of the complex region; the areas
mapped in velocities are within the green square. Right: the velocity field of the ionized gas, obtained with the MPFS at BTA. The position of the supercluster is marked by the cross, the offsets from it are in arc seconds. The contours are the Hα intensities. 1" = 30 pc, North at the top, East at the left.

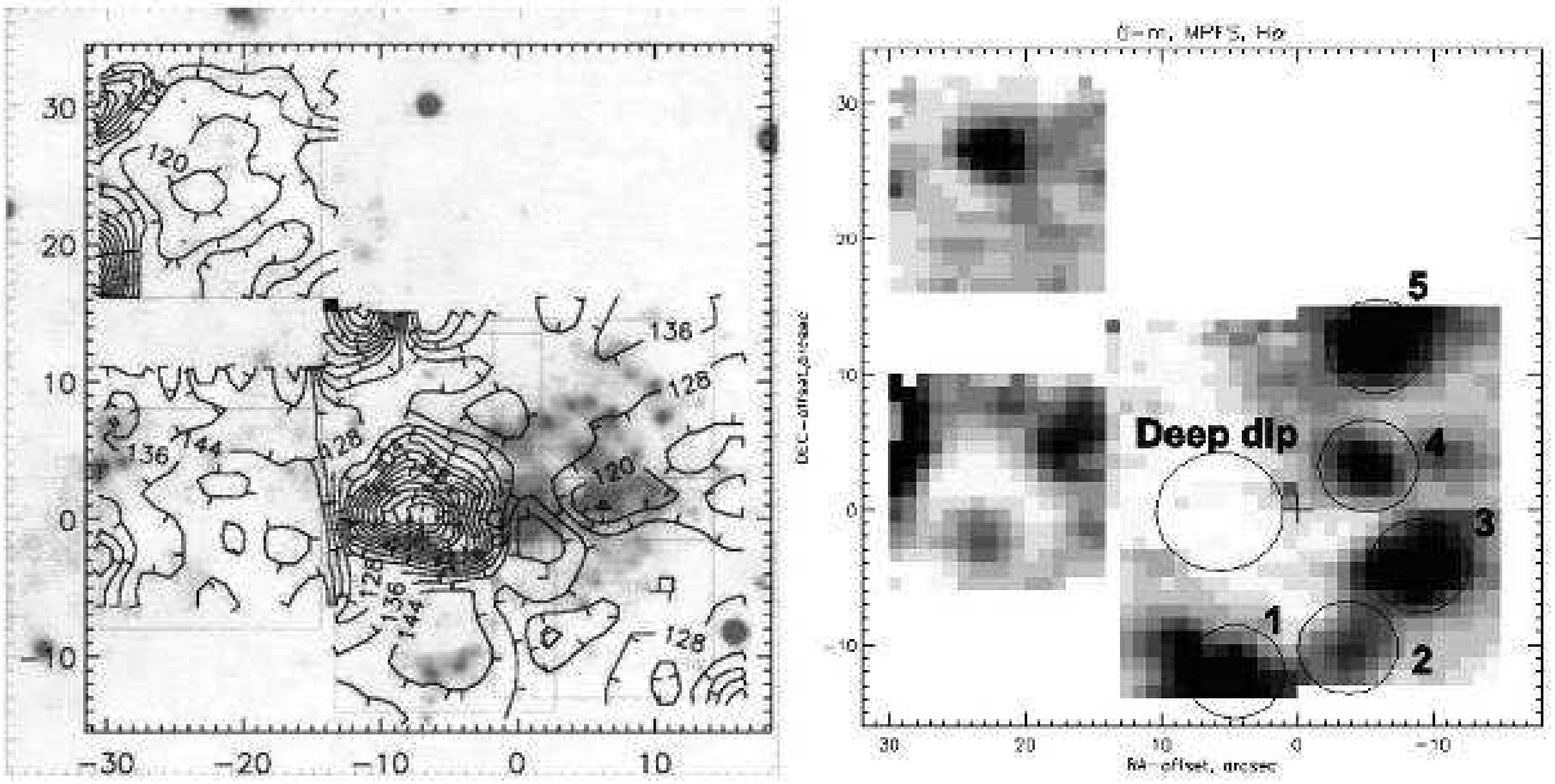

Figure 3. Left: contours of the ionized gas line-of -sight velocity field (BTA MPFS), overlaid on an optical image in R-band (NOT). Right: the Hα map of the same field. The regions studied in more detail are marked (BTA MPFS). The outer north-eastern field is E1 and the eastern field is E2. The coordinates are the same as in Fig. 2, right.

# 3 OBSERVATIONAL RESULTS

## 3.1 The ionized gas data (BTA MPFS)

### 3.1.1 Velocity determinations

The velocity fields inside the complex and two nearby fields were obtained with the BTA MPFS using the emission lines $H_\alpha$, [NII] 6548+6583 and [SII] 6717+6731. These fields are shown in fig. 2 and 3.

Assuming the single velocity values in each position, the errors are about 10 km/s and comparison with the long slit data (figs. 4, 5, 6 in Efremov et al.. 2002) demonstrates a good agreement between both estimations. A roundish depression of 10 x 12" (300 x 360 pc, assuming a 6.0 Mpc distance) in size is seen centered 7" east of the SSC, the radial velocity in its center being about 40 km/s, i.e. ~100 km/s smaller than that of the SSC and the neighboring gas. It is the cross-section of this depression by the HII line-of-sight velocity curve along the PA 83° that was called the "fast expansion" in Efrem ov et al. (2002) - their slit at PA 83° very luckily intercepted this feature a bit to the North of its center. This feature is referred to below as the “deep dip” or the ”crater” (Fig. 3). A smaller and less deep depression is seen NE of the complex outside of it. Both features are within the region of the lowest $H_\alpha$ (and HI, see below) intensity at the star complex’s eastern border.

Two-component Gaussian analysis of emission profiles was carried out for the regions marked in Fig. 3 (right); the resulting velocity distributions are given in Fig. 4 and Table 2.

Table 2. Velocity component values and their dispersions in regions, marked in Fig. 3, km/s

| region | blue Vr | blue FWHM | red Vr | red FWHM |
|---|---|---|---|---|
| Deep dip | 13.0 | 258.6 | 147.2 | 120.4 |
| 1 | 71.8 | 197.2 | 153.2 | 131.5 |
| 2 | 108.2 | 215.1 | 145.6 | 120.9 |
| 3 | 81.6 | 236.9 | 138.8 | 138.3 |
| 4 | 58.4 | 258.8 | 145.4 | 145.6 |
| 5 | 72.9 | 219.8 | 142.7 | 133.6 |
| 1 – 5 | 79±18 | 225±23 | 145±5 | 131±10 |

The observed profiles of emission lines are presented there as the superposition of two Gaussian curves. The values of Vr for each component within the bright regions 1 – 5 in Fig. 4 are rather similar; on average they are +79 +/- 18 km/s (the width is 225+/-23 km/s) and +145 +/- 5 km/s (the width is 131+/-10 km/s). It is worth noting the excellent agreement of the red velocity with the SSC velocity, +141 km/s (Larsen et al. 2006).

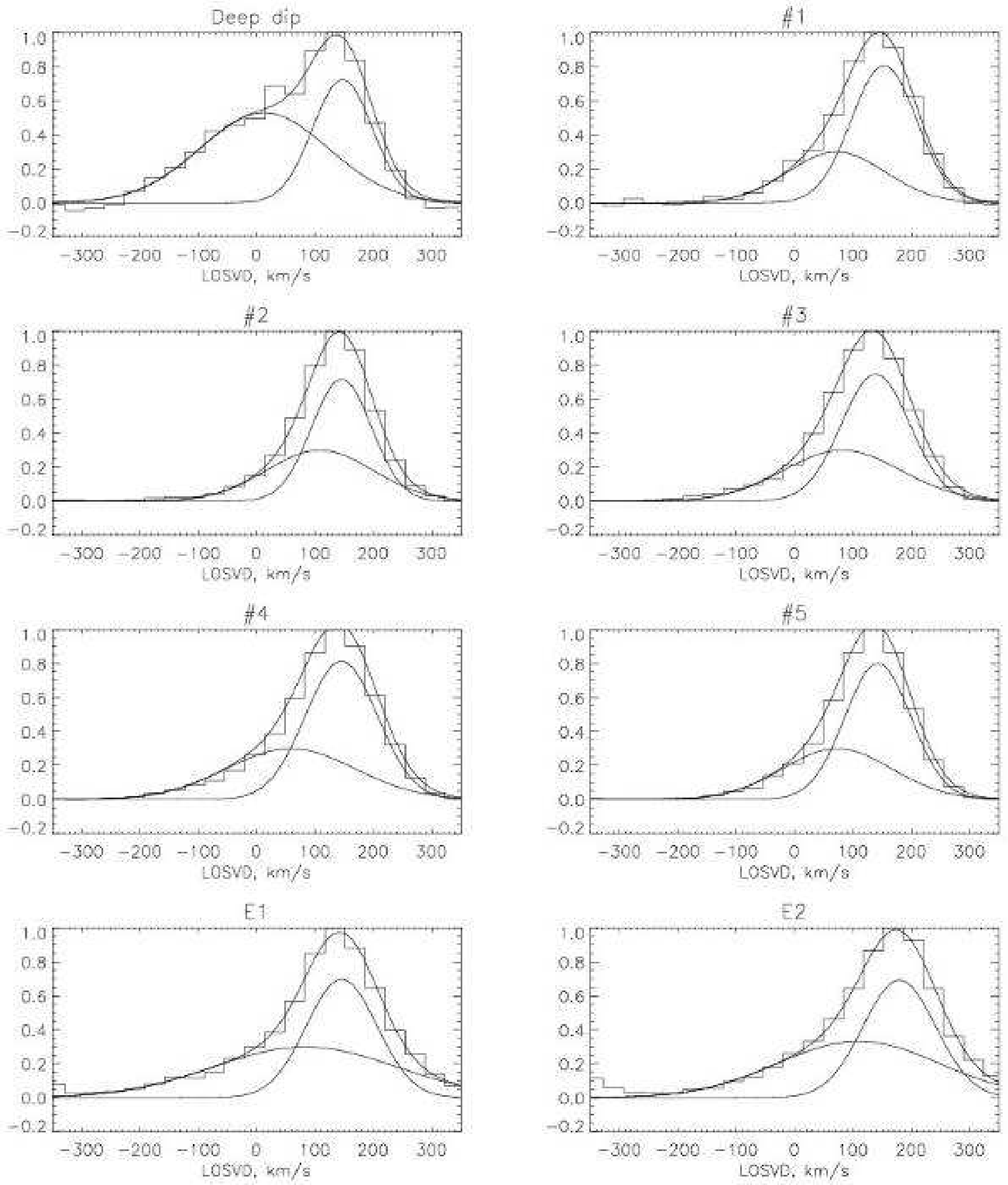

Fig. 4. The line of sight velocity distributions from the emission lines within the deep dip, the regions marked in the Fig. 3 (right) and in the E1 and E2 outer fields. Representations by two Gaussians are given. The velocities mapped in Fig. 2 and 3 correspond to the barycentric (intensity weighted) velocities. See text and Table 2.

Within the dip, the blue-shifted component of the emission lines is wider than in other regions. It corresponds to a velocity of +13 km/s (the width is 260 km/s), whereas the red-shifted component has a velocity of +147 km/s and a width of 120 km/s, similar to values for the regions 1 – 5.

The emission line's barycentric velocities (i.e. intensity weighted, obtained as those Vr values, at both sides of which the integrated values of emission intensities are equal) within all 6 areas are close to those obtained in Efremov et al. (2002) and correspond, of course, to ones shown in our Figs. 2 and 3 above. Taken at face value, the finding of two values of Vr with similar difference inside each area (about 60 km/s), in each one of areas 1 – 5, might imply that all of these areas are rather close to the center of an expanding HII shell with a systemic velocity ~112 km/s and a lower limit expanding velocity of ~30 km/s. However, the red components over the complex are about the same and closely correspond to the SSC velocity and to the average HI velocities near the latter (but outside the deep dip). Note also that (excluding the dip region) the red components are stronger and narrower than the blue ones. The blue components are more different; region #4, closest to the supercluster, has a second (after the dip) blue velocity component. Thus the overall picture looks like a giant semi-shell with a blue-shifted center and an outflow in the same direction. This suggestion is consistent with the HI data, as will be discussed below.

The night-sky Hα line is close to the blue side of the Hα line of NGC 6946 , and, if not properly subtracted, might shift the barycentric velocity to a lower value, especially in regions of low HII gas density. This has led to wrong conclusions in some earlier work on gas kinematics in NGC 6946, as described by Bonnarel et al. (1988). We are careful to exclude this possibility.

The MPFS results were obtained after careful sky line subtraction, based on the spectrum observed far from the galaxy, as was also done by Efremov et all (2002). A cross-correlation of five emission lines was used, including the SII lines, which are completely uncontaminated by sky emission.

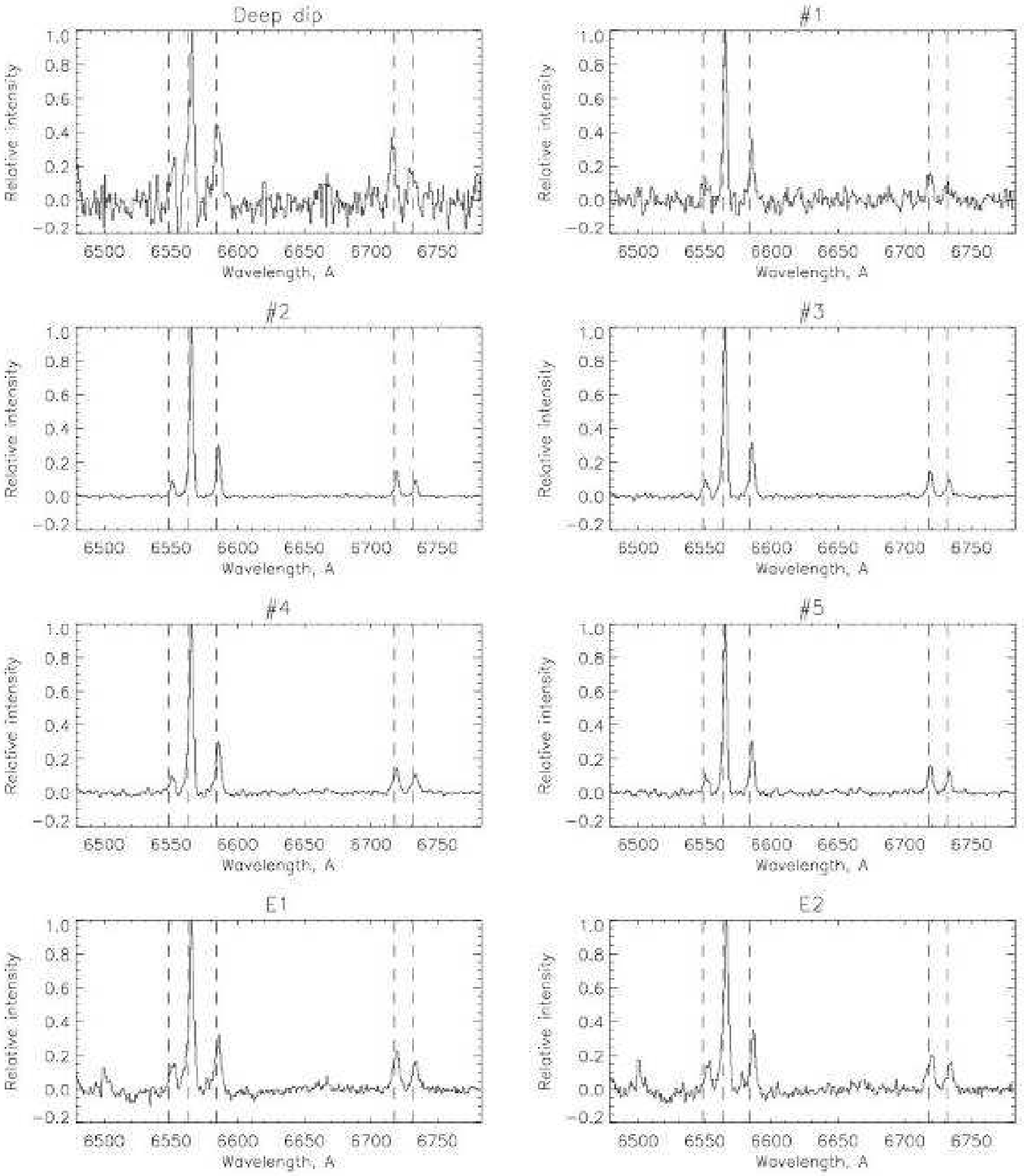

Figure 5. Parts of the BTA MPFS spectra for the deep dip, the regions 1– 5 and the outer regions E1 and E2. The laboratory wavelengths of the emission lines Hα λ 6562 Ǻ, [NII] λ6548, 6584 Ǻ and [SII] λ6716, 6730 Ǻ are shown by vertical lines.

### 3.1.2 Mechanism of ionization

From the ratio of emission line intensities within the deep dip ("fast expansion") Efremov et al. (2002) noted signatures of shock excitation in that region. With the BTA MPFS data we are able to reinforce this conclusion strongly and to obtain quantitative estimations. Fig. 3 (right) is a monochromatic map of the complex in the $H_\alpha$ line, where the regions are marked within which the emission lines were studied in detail. It is seen that the deep dip is near the center of a vast area in which the HII emission is quite low.

In Table 3 and Fig. 6 the ratios of intensities of emission lines in these areas are given for [OIII] and [NII] emission lines relative to $H_\alpha$ and $H_\beta$.

Table 3. Emission line parameters inside the deep dip and within HII regions

| region | Offset RA | Offset DEC | Log(I([NII])/I(Hα)) | Log(I([SII])/ I(Hα)) | Log(I([OIII])/ I(Hβ)) |
|---|---|---|---|---|---|
| 1 | -5” | -12” | -0.365 ± 0.013 | -0.531 ± 0.039 | -0.563 ± 0.111 |
| 2 | +3 | -11 | -0.292 ± 0.054 | -0.422 ± 0.064 | -0.462 ± 0.142 |
| 3 | +9 | -4 | -0.338 ± 0.019 | -0.552 ± 0.028 | -0.547 ± 0.056 |
| 4 | +5 | +3 | -0.308 ± 0.022 | -0.521 ± 0.066 | -0.322 ± 0.152 |
| 5 | +6 | +12 | -0.351 ± 0.030 | -0.560 ± 0.041 | -0.454 ± 0.109 |
| Deep dip | -6 | 0 | -0.003 ± 0.120 | 0.062 ± 0.108 | -0.140 ± 0.100 |

The dividing line between photo-ionization due to the presence of hot stars, or ionization produced by shock waves, is presented in Fig. 6 according to the diagnostic tool, developed by Veilleux and Osterbrock (1987).

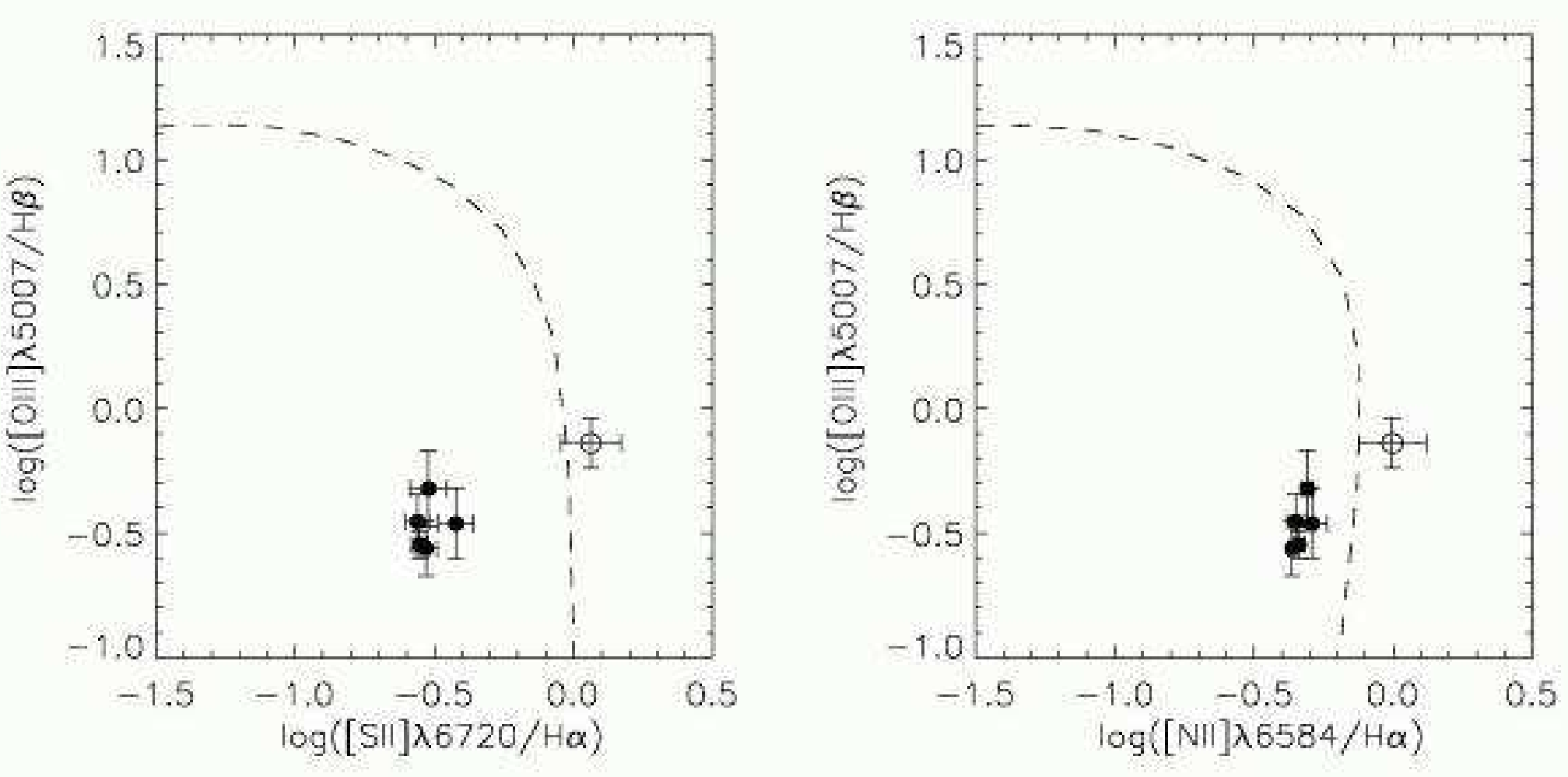

Figure 6. The diagnostic diagram for the emission line ratios (MPFS) in the regions marked in Fig. 3 (right). All ratios, except those for the deep dip region (shown as open circle), correspond to the ionization from hot stars, being to the left of the dividing curve.

The line intensities and their ratio maps are presented in Fig. 7, where we can see how the low gas density (HII, NII and SII) is associated with the region of ionization by shocks. It is seen also that ionization by shocks appears to be only present within the velocity depressions - the deep dip and a small region north of it.

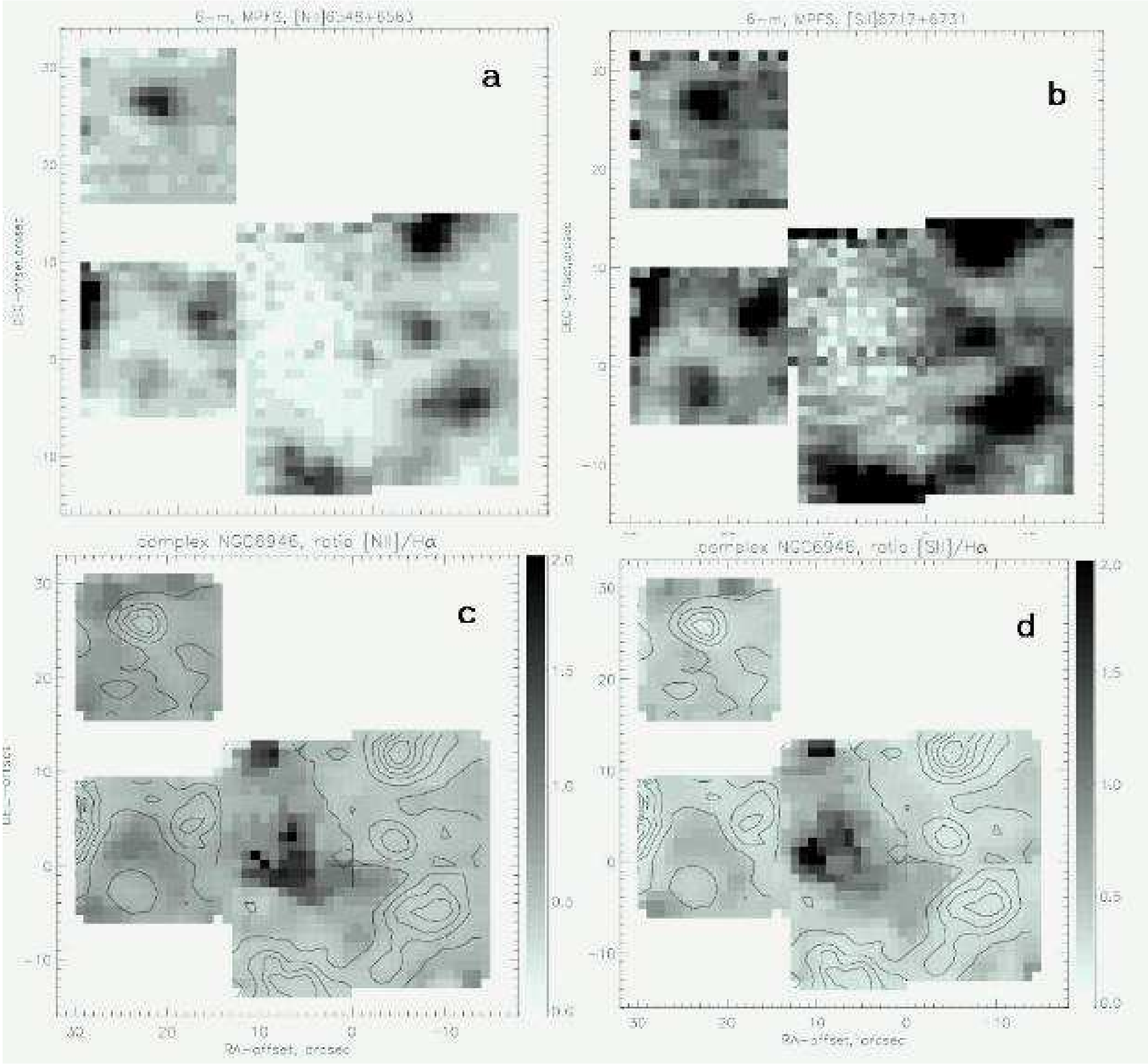

Figure 7. Maps of the emission line intensities and ratios (MPFS): a) [NII] 6548 + 6583, b) [SII] 6717 + 6731, c) [NII]/Hα, d) [SII]/Hα. The contours are Hα intensities. Compare with Fig. 2.

The low density of all gas species within the deep dip region leads to a lower signal/noise ratio for emission line intensities there. Because of this, no conclusion was obtained for this region with the WHT INTEGRAL IFU data (see below). However, as shown in Fig. 5 and Table 3, the BTA data for the line intensities within the deep dip present good enough S/N for diagnostic analysis. It is seen that the [NII]/$H_\alpha$ ratio within the deep dip is indeed larger than in all bright emission regions 1 – 5.

The parameters of emission lines were also measured with the WHT in the same bright HII regions as ones measured with the BTA and were found to conform to expected ratios from photo-ionization (see below in Section 3.3).

It is worth noting that in the galaxy NGC 1084 (Moiseev, 2000) an increase of the [NII]/Hα ratio in the regions of ionized gas deviating from circular rotation was explained as a signature of shock wave fronts in these regions.

### 3.2 Velocity slice through the deep dip (Gemini North)

The issue of the radial velocities within the deep dip is crucial to understand this feature. The relevant data were obtained by N. Bastian at the Gemini North telescope with the GMOS-N spectrograph on the night of 2006-05-25. The slit-width was set to 1", using the grating R831 centered around λ 6600Å . This configuration provides a plate scale of 0.14"/pix and a dispersion of 1.3 Å/pix, with a resolution of 3A around Ha. The total integration time was 900 s. The spectral range was 5550 – 7600 Å. No calibration in flux was done.

The data were reduced using the standard Gemini reduction pipeline.
First, only the Hα line profile was used, fitted with a single Gaussian, sky line subtraction being done from the outer part of the same long slit spectrum, outside the main body of the galaxy. The slit position is shown in Fig. 8; the resulting line-of-sight velocity and flux profiles are in Fig. 9.

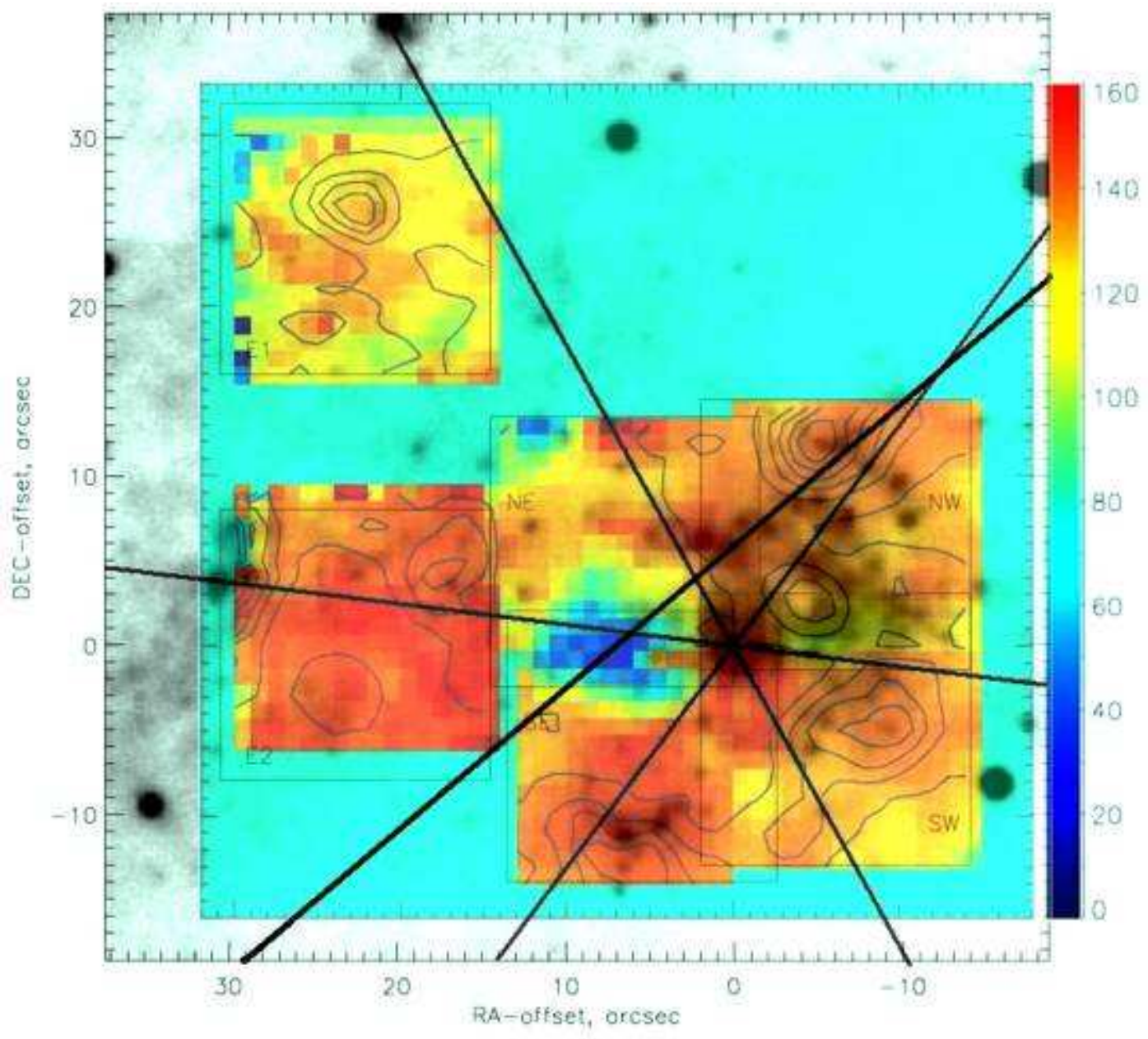

Figure 8. The slit positions of BTA (used in Efremov et al. 2002) and Gemini North, overlaid onto the velocity field and R-band image. The Gemini North slit goes through the deep dip at PA -50°.

We have also attempted to get line-of–sight velocities from the lines of [NII] 6584 and [SII] 6717, 6731, which are well separated from any significant telluric lines for the range of velocities observed in the target spectrum. Thus, we lose emission signal in favor of a better separation in the spectral axis and in favor of being completely free from any sky line contamination.

In order to increase the signal to noise ratio of the 1D-spectra, we have binned three rows along the spatial axis, leaving unaltered the spectral axis. Line centroids, intensity peaks, line fluxes, FWHMs, and their corresponding errors, have been calculated making use of the DIPSO program in the STARLINK package. This analysis has been applied to the three emission lines mentioned above, as well as to two significant telluric lines within the analyzed spectral range. The details of reduction will be published elsewhere (Sánchez-Gil et al., in preparation).

A weighted mean radial velocity curve has been calculated from the radial velocities of the individual lines weighted by the inverse of the radial velocity errors. The uncertainty of the mean radial velocity has been also calculated following classical statistical procedures. In the same way, the mean velocity dispersion and its error have been estimated from the same data (Fig. 10).

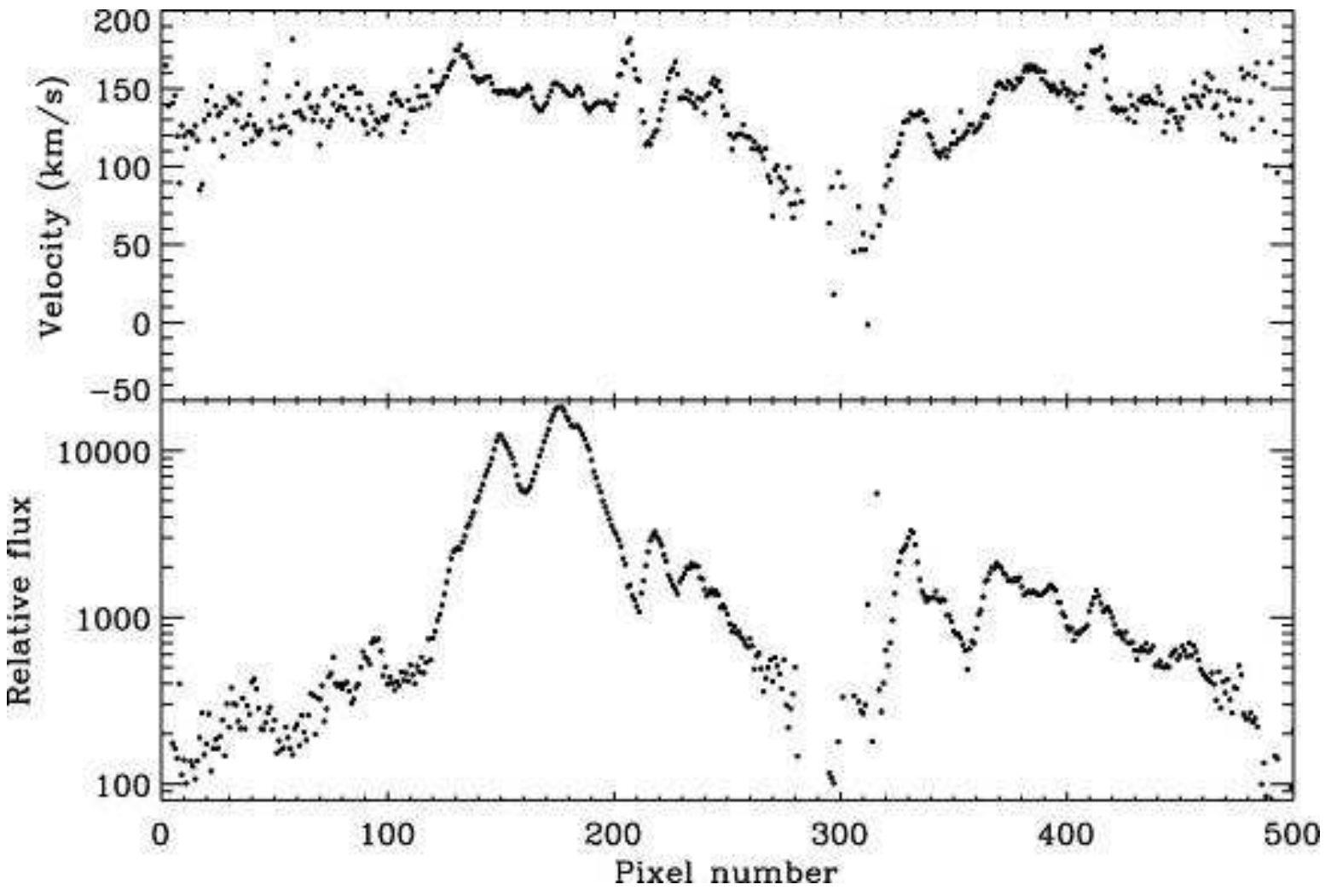

Figure 9. The data obtained from Hα with the Gemini North telescope.
Top: the velocity curve, bottom: the flux profile. 1 pix = 0.14”. SE at the right, NW at the left.

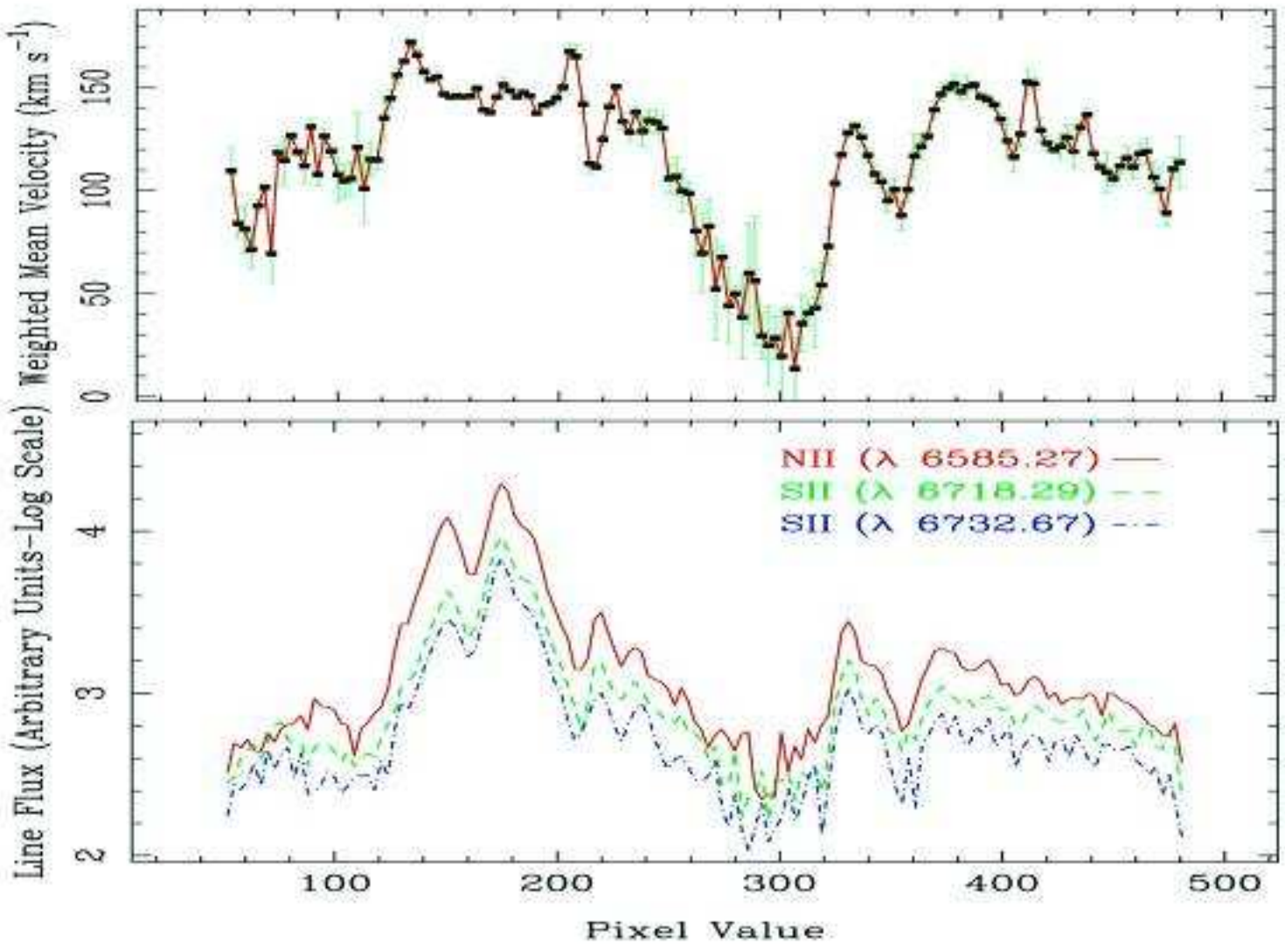

Figure 10. The data obtained with the Gemini North telescope from the [NII] and [SII] lines. The pixel numbers are the same as in Fig. 9.

It is important to note that the radial velocity curve derived from forbidden lines (Fig. 10) agrees well with that obtained from the Hα line (Fig. 9). The somewhat lower velocities that are found from the forbidden lines within the regions of low flux may indicate a lower velocity of the shocked clouds, like it was found in NGC 1084 galaxy (Moiseev, 2000). There is also good agreement with velocities obtained from the MPFS velocity slice along the Gemini slit position (PA -50º). Thus we may affirm that our velocities are reliable even in the regions of the lowest density of ionized gas.

### 3.3 The western rim

The western edge of the star/cluster complex is unusually sharp and lies along a circular arc 130-degree long (Fig. 1, 13, and 20). Such a regular stellar structural shape, to the best of our knowledge, is otherwise known only for the giant stellar arcs in the LMC (Hodge, 1967; Efremov, 2004b,c) and for a few galaxies which appear to have been in interaction with the IGM, such as NGC 2276 and DDO 265. In the latter cases the arc-like shape is evidently due to star formation, triggered in a bow shock by ram pressure along the galaxy's leading edge. In a few other galaxies an arc-like (bow shock) shape is observed in the outer HI density contours (Efremov, 2002 and ref. therein).

The issue arises whether this sharp regular edge is due to a large dust density along the western rim or it is the real shape of the star complex. In a map of extinction over the whole of NGC 6946, constructed from B and K-band photometry, by Trewhella (1998), an arc of high extinction around the complex's west edge may be noted, with $A_B$ being mostly between ~1.0-2.0, whereas it is low in the central part of the complex. Similar but more detailed results were obtained from the HST stellar UBV photometry within the complex (Larsen et al. (2002).

The complex was observed in August 2003 by E.Alfaro and B.García-Lorenzo with the INTEGRAL device at the WHT telescope. Within the bright HII regions located at the western and southern parts of the complex, the light absorption based on the $H_\alpha/H_\beta$ ratios was determined and translated into magnitudes of B-band absorption; the resulting map is given in Fig. 11.

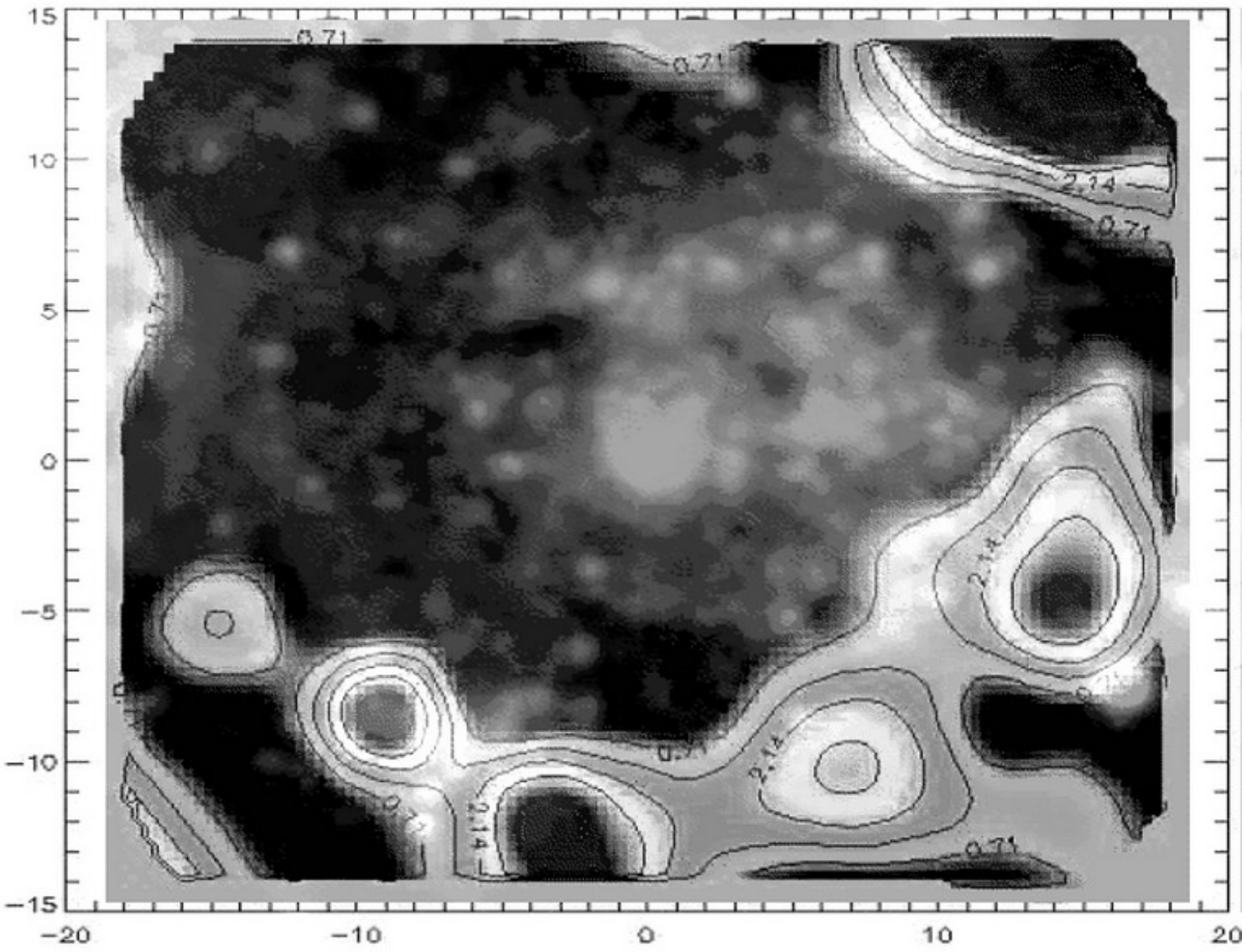

Figure 11. Contours of $A_B$ light absorption based on WHT data. Values range from < $0.71^m$ over most the complex to > $2.85^m$ at the upper right. The contours are overlaid onto the image from Fig. 2 left. The X,Y axes are in arc seconds, the supercluster is at (0, 0) and the deep dip centre is at (-7, 0).

It is seen that inside the bright HII regions surrounding the star complex at the West, $A_B$ values are as large as $3.5^m$ , whereas inside the complex $A_B$ is smaller than $0.7^m$. The bright HII region 4 (Fig. 3) is inside of the complex and is unseen in Fig. 11.

The high light absorption along the complex's western edge is confirmed by Fig. 12, where the K-band image obtained by S.Larsen with the Keck-II telescope is given. The optical depth in the K-band being 10 times smaller than in the B band, the objects behind the dust clouds may be seen and indeed, in Fig. 12 the complex has neither a sharp nor a regular edge at the west or anywhere else. These results suggest that the arc-like western rim is mostly outlined by high light extinction, as was proposed earlier (Elmegreen, Efremov & Larsen, 2000).

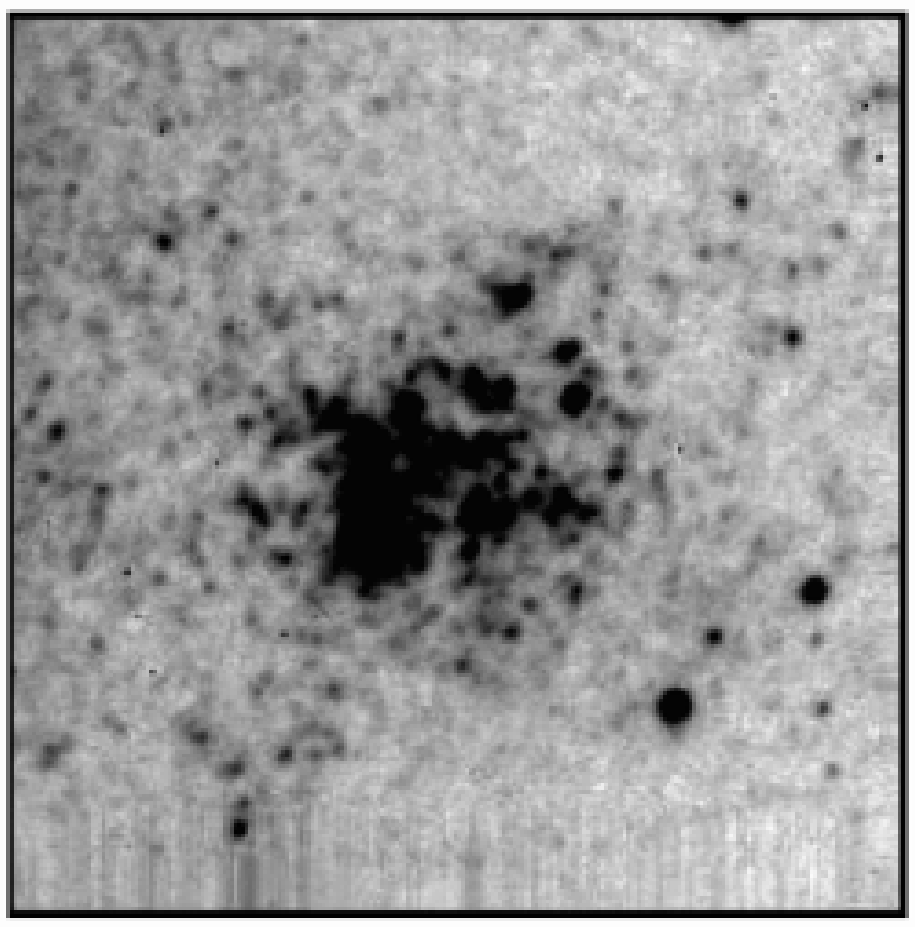

Figure 12. The Hodge complex in K-band (the Keck-II image). The image was taken with the slit-viewing camera (SCAM) of the Nirspec spectrograph and is a sum of 3 exposures of 60 s each. The pixel scale is 0.183"/pixel and the detector has 256x256 pixels, so the total field
is 47” across. North is at top.

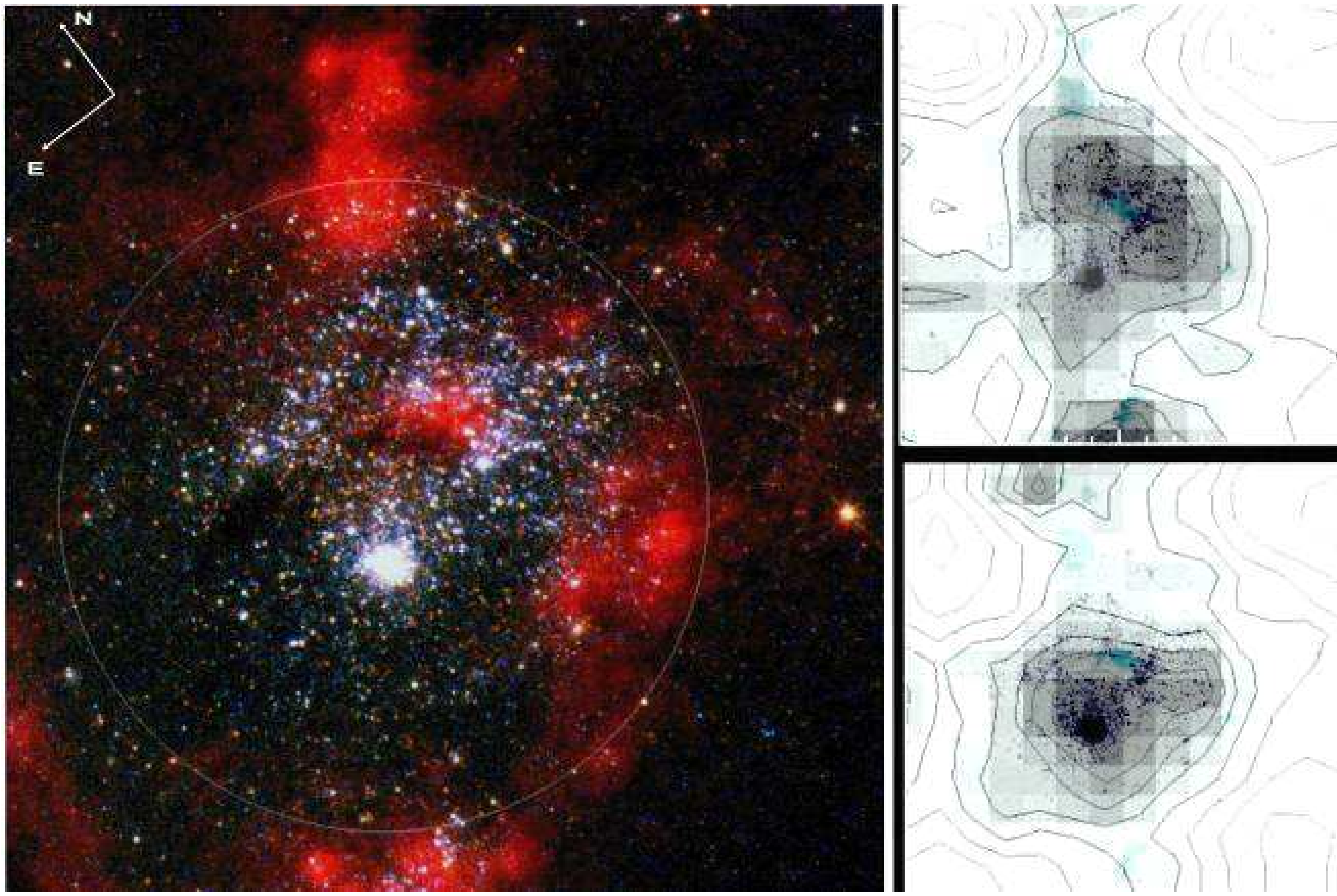

Figure 13 . Left: the HST PC image from Larsen et al. (2002) with a circumcomplex circle added. Upper right: the contours of the bright main sequence (blue) star densities. Bottom right: the contours of the red supergiant densities. The contours were taken from fig. 9 in Larsen et al. (2002) and overlaid onto the negative complex image.

The older red supergiants are concentrated to a clump between the western rim centre and the SSC, as is seen by eye (Fig. 13 left) and by star density averaging (Fig. 13 right) in the HST data. This older population might have triggered the formation of a younger one, which would be compatible with the results from Larsen et al. (2006), who found a super-solar [alpha/Fe] ratio, usually interpreted as a result of rapid, bursty star formation from gas enriched by a massive SNII. Indeed, the SF history inside the complex provides evidence for a maximum rate of star formation at least 10-15 Myr prior to the formation of the SSC (Larsen et al. 2002).

At any rate, the spatial pattern formed by circular arrays of different aged stars, observed in Fig. 13 (see also Fig. 20 below), is difficult to explain. What is still surprising is that the western rim of the complex, the circular arc, formed or not by dust, is concentric to the arc of the bright blue stars inside the complex, the opening angle of both arcs being similar. Also, the arcs of clusters are seen not only along the western rim, but (shorter ones) at the complex east too, again with the same centre, which is also the centre of the red supergiant concentration. It is the red supergiant clump - and not the SSC - which is the core of the Hodge complex.

### 3.4 The HI gas data (WSRT)

Relevant data for our region were obtained by R.Boomsma while studying the distribution of HI over all of NGC 6946 using the Westerbork Synthesis Radio Telescope (WSRT). The results for all the galaxy are given elsewhere (Boomsma, 2007; Boomsma et al., in preparation). It was found that the Hodge complex is inside a region of low HI density, but an oval region of even lower HI density (a hole) is seen to the east and northeast part of the complex (Fig. 14) , being remarkably similar in position and shape to the region of the shock excitation and low ionized gas density, shown in the Fig. 7.

This HI hole is #85 in the list of HI holes in NGC 6946 (Boomsma, 2007); its size given there is 1.4 x 1.0 kpc. The hole is centered on the deep dip position and not the supercluster, the latter being at the western edge of the oval region of the lowest HI and HII density. The complex is surrounded with HII regions (Fig. 14 and 15).

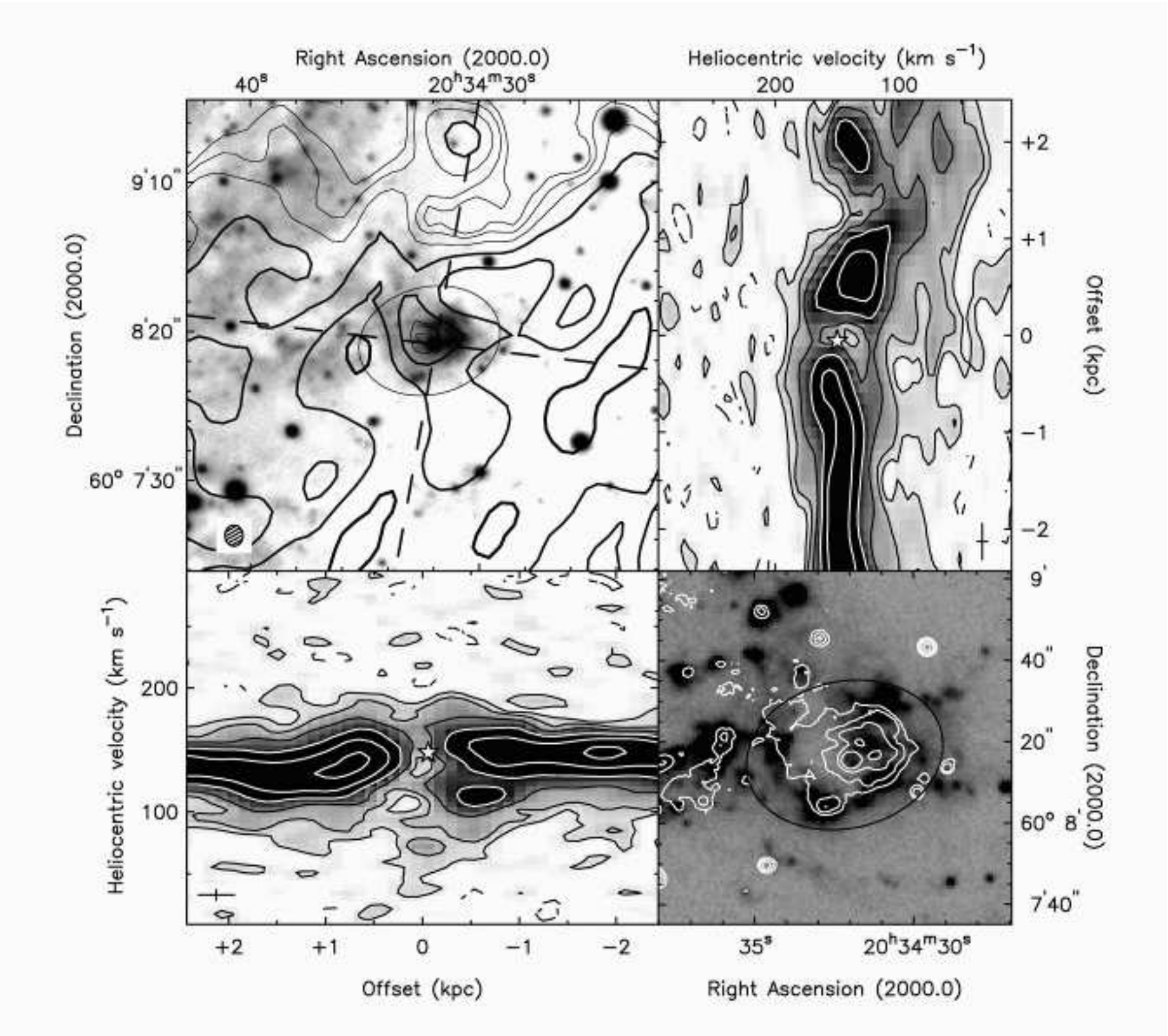

Figure 14. The WSRT data. Top left: B-band image with HI density contours at Vr = 150 km/s. The dashed lines show the positions of the slices at PAs -10° and 83°, the ellipse indicates the extent of HI hole and the shaded ellipse indicates the beam. Top right: the position-velocity diagram at PA -10°, the bottom left – the position-velocity diagram at PA 83°, both on the same scale as the top left image. The contour levels are at -0.24, -0.17, 0.34, 0.68, 1.4, 2.7 and 5.4 mJy/beam. The stars show the position and velocity (adopted to be 141 km/s) of the SSC. The crosses indicate the resolution. The bottom right: a larger scale image of HII regions (dark) with the B-band contours overlaid. The possible HII supershell is shown as oval.

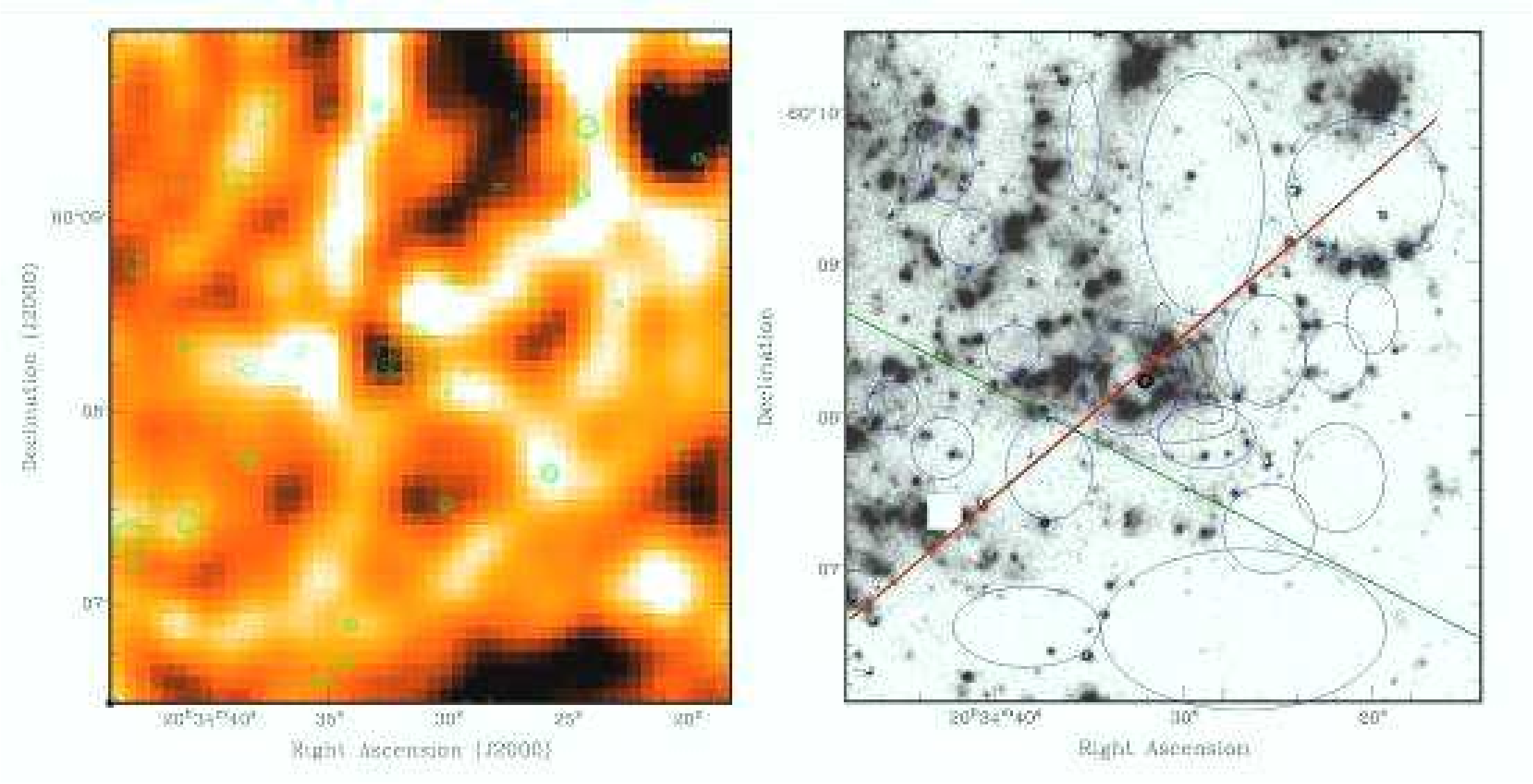

Figure 15. Left: the density map of the neutral hydrogen in the field of the Hodge complex (WSRT data). The low density regions are dark, the contours are for the R-band optical image. Right: the Hα map (from Fergusson, Gallagher, Wyse 1998) with the positions of the Gemini slit (red, PA -50°), the galaxy major axis (green) and the su percluster (small circle) added. The contours at the complex West are for HI densities around velocity 110 km/s (see Fig. 14), the ovals are approximate positions of HI holes, found in Boomsma (2007).

Figs. 16 and 17 compare the intensities and velocities of the HII and HI gas in the wide area surrounding the complex, demonstrating that the overall velocity of the HII and HI gas in the complex appears to be in agreement with the velocity of the surrounding regions, determined mainly by the galaxy rotation.

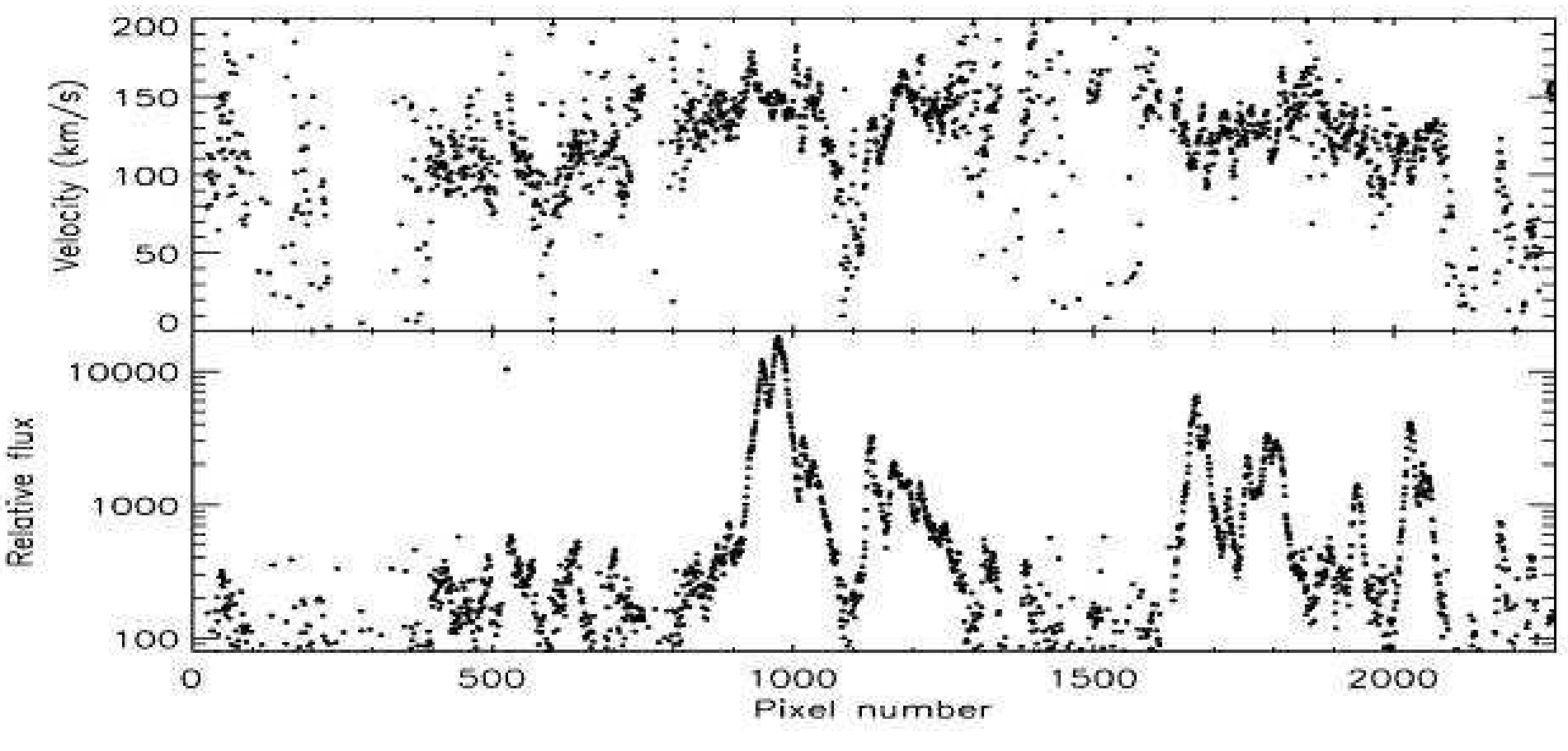

Figure 16. The full slit length (5') data obtained for velocity (top) and flux (bottom) of the Hα line from the Gemini North telescope. SE at right, NW at left. 1 pix = 0.14". The deep dip is seen near pixel 1100.

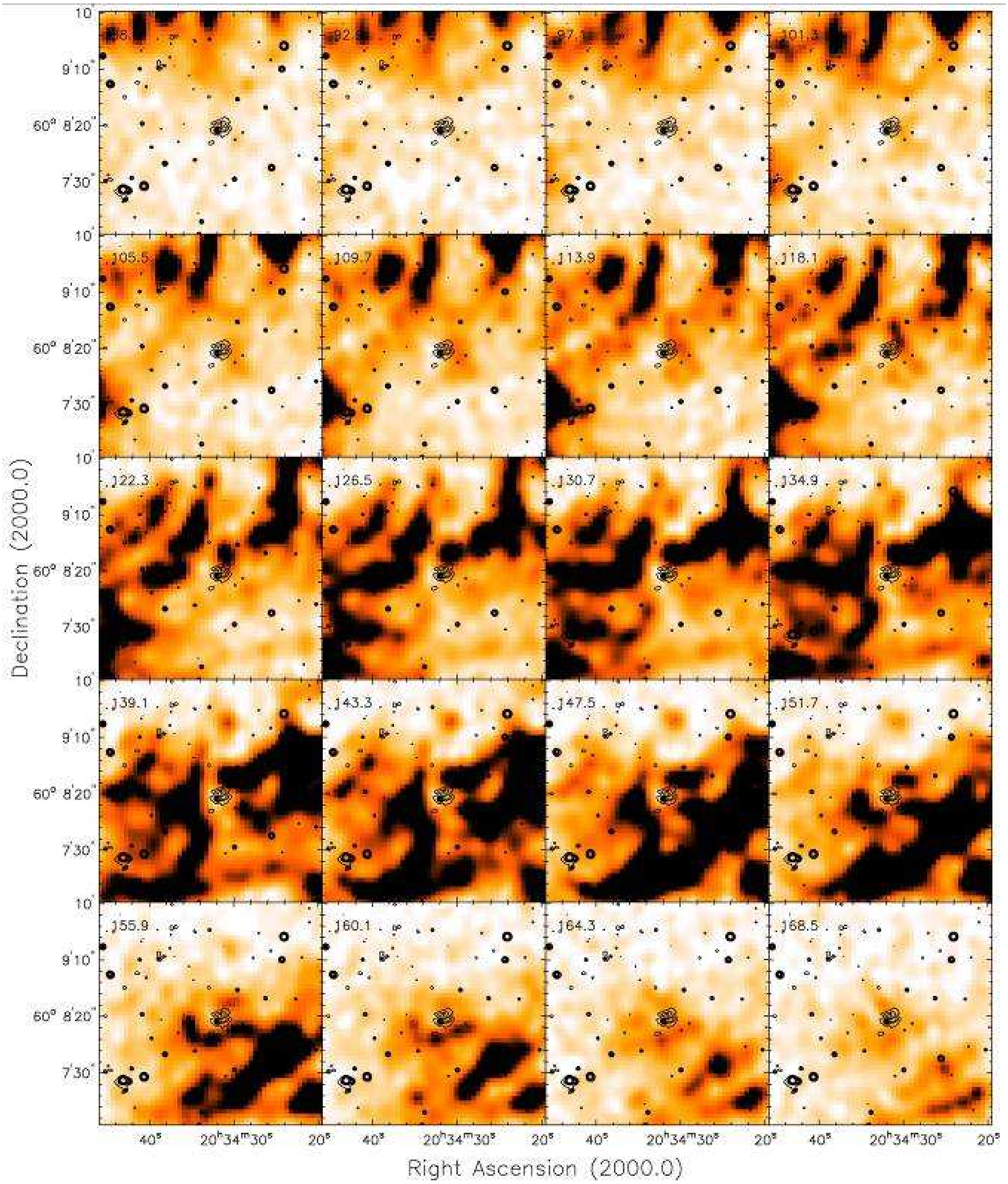

Figure 17. The channel maps of HI density for the Hodge complex region. The contours of the R-band optical image are shown. Velocities are given at the left top. Compare with Figs. 14 and 15. WSRT data.

### 3.5 Gas velocity fields

#### 3.5.1 Circumcomplex shell expansion

We are now in position to compare the results obtained for the velocities of the ionized and neutral gas in the field of the complex. The velocity fields of HI and the ionized gas are overlaid in Fig. 18. The agreement is mostly poor, as might be expected. As we dealt with the intensity-weighted velocity fields, the locations of velocity extremes in the sky plane is a result from the interplay of respective intensities and velocities. The rather low HI resolution (13” comparing with 1’’ in HII) explains why the most remarkable feature of the ionized gas velocity field, the deep dip, is unseen in HI velocity map, though it is observed in the HI velocity slices.

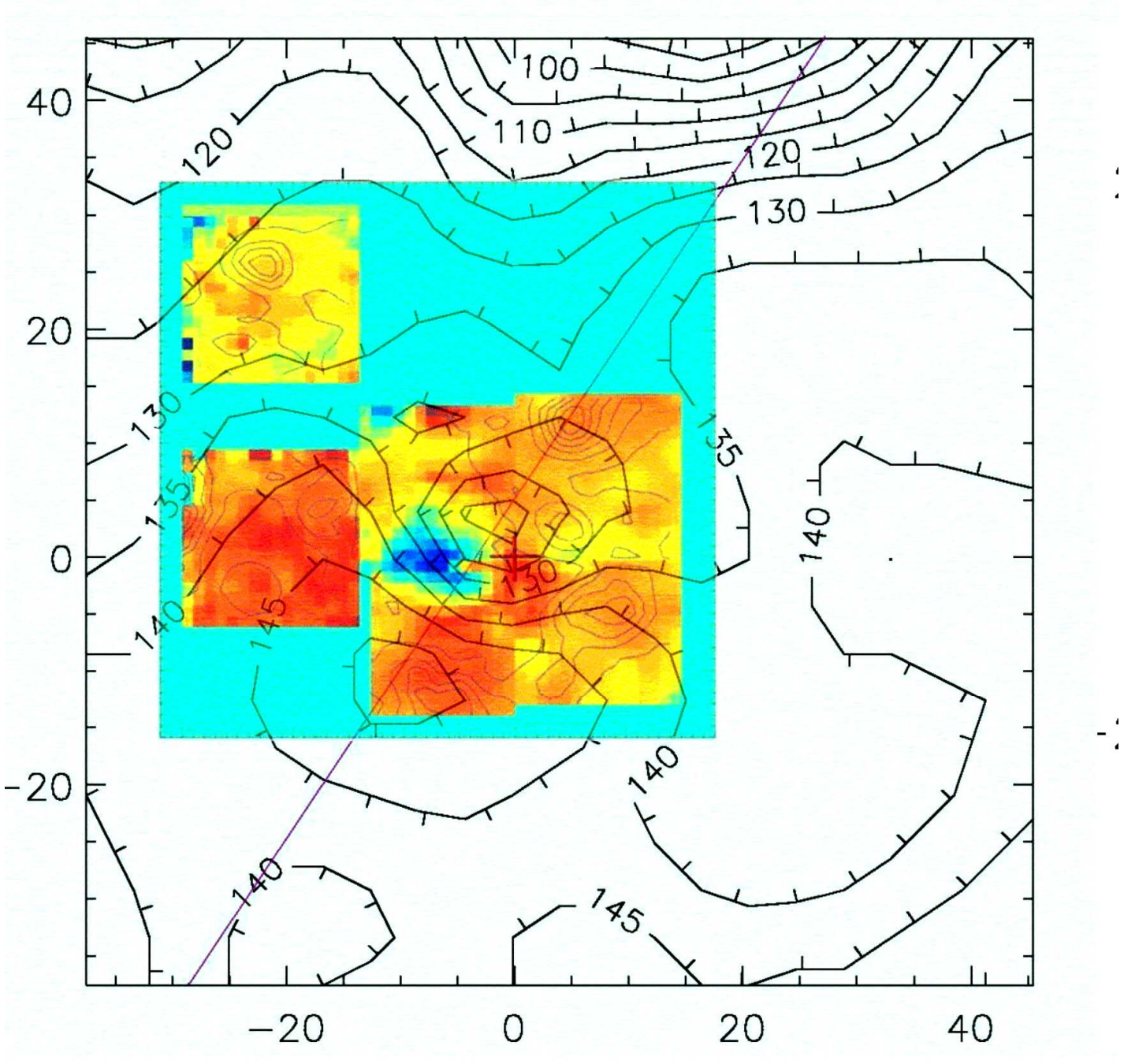

Figure 18. The velocity fields for HI (WSRT, the ticked contours) and ionized gas (BTA, colored areas, see Fig. 2, right). The position of the supercluster (at 0,0) is marked by the cross, the deep dip is the blue spot. The line of the largest gradient in the HI velocity field is shown, its PA= -30°. The conti nuous faint contours in the insert are for the HII emission intensity. The HI Vr contour lines have tics to the lower velocity side.

The line going through the local HI velocity extremes in Fig. 18 (left), has the position angle PA -30°, which is close to the orientation o f the galaxy kinematical minor axis: - 32° from HII data (Bonnarel et al., 1988) and -38° fro m HI (Boomsma, 2007). A similar velocity gradient is seen also in the ionized gas velocity curve for PA -37°, at a distance -20” -- +10” from the SSC ( Efremov et al., 2002, fig. 6 bottom), and it may be suspected also in the velocity field of the ionized gas. (Otherwise the respective features in the latter might be a part of the Hα velocity bump, surrounding the dip; see Fig. 19).

The orientation of the line of maximum velocity gradient close to that of the galaxy minor axis is a signature of either expansion or contraction within the galaxy plane (see for example fig. 5 in Begum and Chengalur, 2003). Considering that the near side of NGC 6946 is in the NW and the Vr local minima is also at that side (Fig. 18), the conclusion is that we are dealing with expansion. The position of the center of expansion (the point at the line of maxima gradient equally distanced from maxima (150 km/s) and minima (120 km/s) closed contours of Vr) has the intermediate Vr = 135 km/s; the center is closer to the deep dip than to the SSC (Fig. 18). The velocity of the SSC is quite close to the median value between these extreme velocities; according to the recent determination by Larsen et al. (2006) it is 141 +/- 2 km/s, based on Keck IR spectra of the red supergiants in the SSC.

The extreme difference of HI velocities along the line of maxima gradient is 30 km/s and, assuming the galaxy gas disk inclination is 38°, this $V_r$ gradient might point to expansion of the HI gas in the plane of the galaxy with a velocity ~30 km/s.

This value is the same as the one suggested by the splitting of emission lines in the bright HII regions 1—5 (see above, Section 3.1). This evidently implies that the HI and HII supershells expand with the same velocity. Note that the higher (redder) velocity of HI is close to the local rotational velocity of the galaxy, whereas the smaller (blue) is much lower. The same is true for the ionized gas velocities.

Large negative deviations are seen in velocity slices of both HI and ionized gas in the deep dip position and around it (Fig. 19).

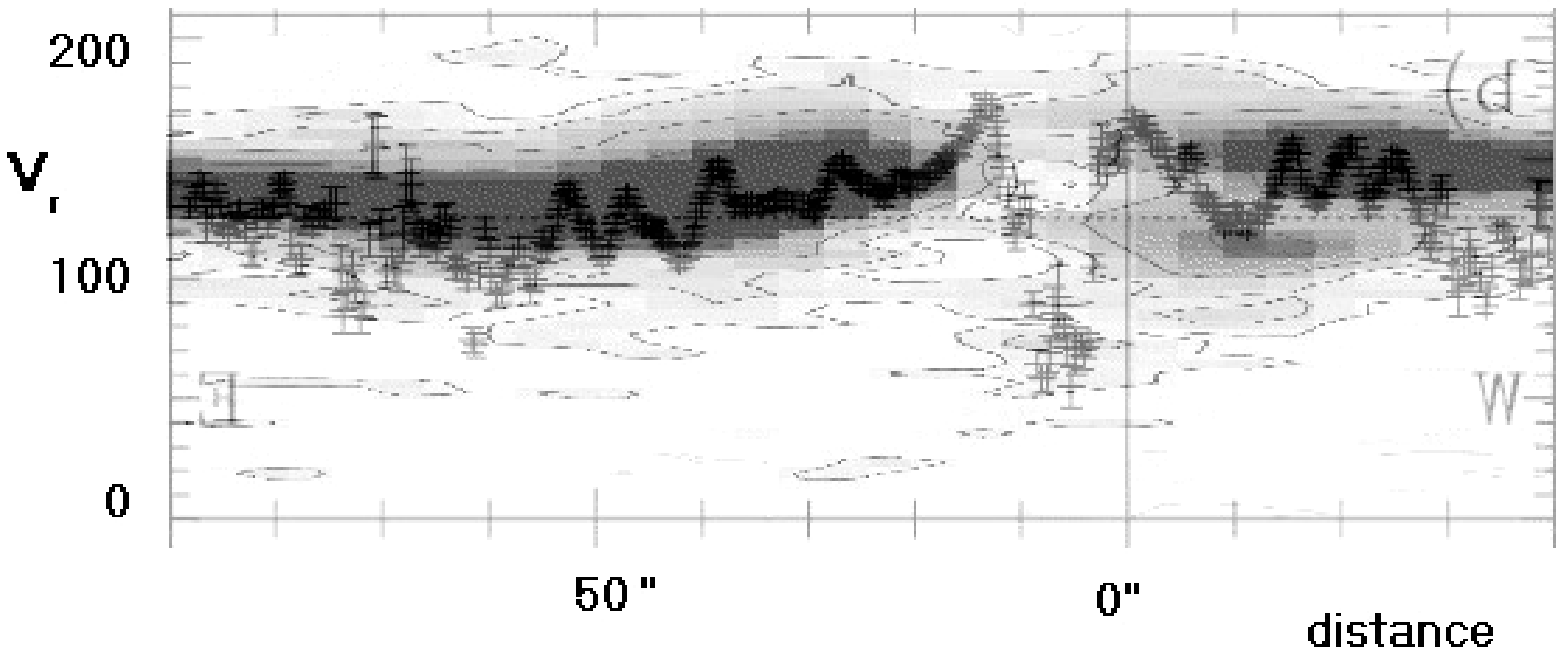

Figure 19. The HI and Hα velocity curves along the PA 83°, going through th e SSC and the deep dip positions. HI (WSRT data, wide strip) and Hα (data from fig. 4b in Efremov et al. 2002, crosses) velocities curves demonstrate the crater-like appearance. East to the left, 0 is at the SSC position.

The features of these slices are in general agreement with the double values of velocities of the ionized gas, found in Section 3.1 within the deep dip and the bright HII regions. Also, the HI holes or loops are seen in each of the HI slices through the SSC at locations of HI density minimum (Fig. 21 below).

We may conclude that the overall agreement of the neutral and ionized gas line-of-sight velocities suggests the expansion of a circumcomplex shell with velocity ~30 km/s, although the picture is surely more complicated. The red velocities of both ionized and neutral gas are rather similar in different positions , whereas the blue ones are quite different.

The largest degree of shock ionization observed just at the deep dip position (Fig. 7) might imply that we observed there the shocked remnants of impacting clouds which have passed through the galaxy gas disk and have triggered the expansion of the supershell.

### 3.5.2 One-sided outflow from the circumcluster shell?

However, another explanation is possible for the nature of the deep dip. It may be the manifestation of a fountain of gas that was blown out from the galaxy disk by the pressure from the hot stars and the SNe of the supercluster.

Earlier the pressure from the SSC and the non-uniform density in the ISM within the complex were suggested (Efremov et al. 2002) to explain the "fast expansion" (called here the deep dip) and its position far from the SSC. These authors found also a shell of HII gas around the supercluster with radius ~6" (the circumcluster shell). Now we note that the distance between the centers of the dip and the supercluster is the same 6 or 7". This suggests that the deep dip is nothing but a manifestation of the blow- out of the circumcluster shell from the (inclined) complex gas plane.

In this case we have not two shells, the fast expansion and the circumcluster shell, (apart from the encompassing HII/HI supershell), but we observe (as the fast expansion) the break-through of the single circumcluster shell. The outflow is one-sided because the supercluster is closer to the near side of the gas layer, the distance of the supercluster to this effective edge of the gas disc layer being a bit smaller than the present day radius of the circumcluster shell. The break site is located rather far (at 7") from the supercluster because the gas disc of the complex is inclined to the sky plane. We may be observing a classical chimney, produced by the energy output from the supercluster, as was found in the NGC 1705 dwarf galaxy, which also hosts a supercluster (Meurer et al., 1992).

The bump in the Hα and HI velocity curves, at the top of which is the dip (Fig. 19), might then be explained by the returning motions of the gas clouds, which were blown out through the chimney earlier. These clouds are naturally expected to be located more or less symmetrically around the chimney axis. The latter, being orthogonal to the complex gas plane, is inclined to the line of sight and this may explain the asymmetrical appearance of the velocity dip walls in Figs. 2 and 9. The shock waves in the blown-out region explain the shock ionization in this region. (Note, anyway, that all these features are seen also in some models of a HVC impact (Santillán et al. 1999).

The outflow hypothesis seems to have a confirmation. As is seen in Fig. 20, the complex is bordered by a nearly regular circle, in spite of the galaxy plane inclination being 38° (Boomsma 2007). We cannot believe the star complex is a spherical feature (the complex diameter is larger than the thickness of the young stellar disc), and at this inclination a circle within the galaxy plane should look like an ellipse, the minor axis tips of which should be at the ticks imaged in Fig. 20.

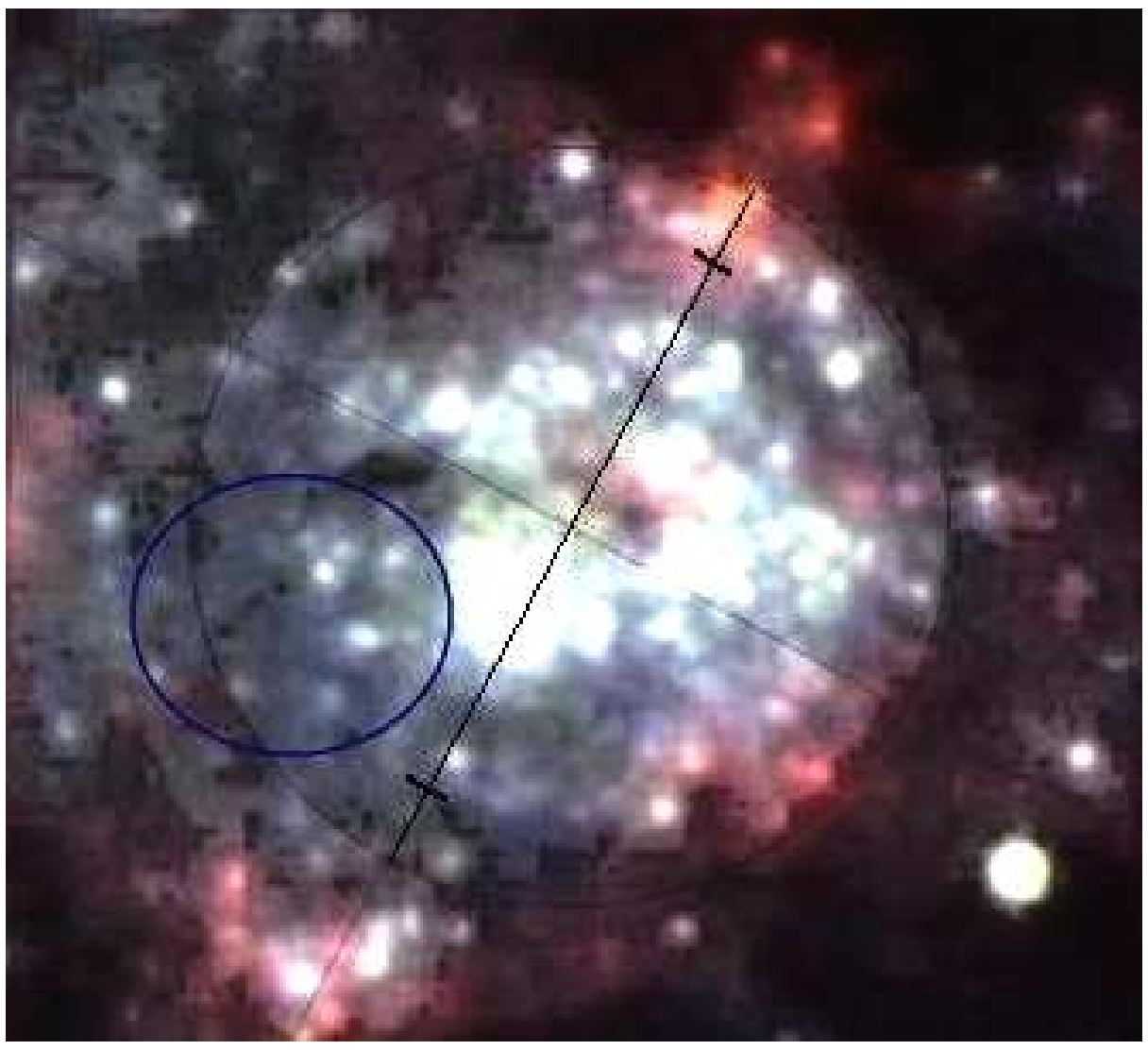

Figure 20. The cenhanced image of the complex . The circle around the complex has the diameter 22" = 660 pc at 5.9 Mpc (Karachentsev et al. 2000). The position angle of major axis is 62°, minor axis -2 8°, the galaxy inclination is 38° (Boomsma 2007). The position of the deep dip is shown by the oval. See text.

In order to see the observed circular edge, the complex plane should be tilted by some 40° in respect to the main plane of the galaxy, having the node line almost parallel to the minor axis of the galaxy. Otherwise, it may be the local corrugation of the galaxy plane in the complex region. These corrugations are the common features of galactic discs, including ours (Alfaro et al. 2001).

Due to the complex plane inclination we do not see the site of the gas break-through from the circumcluster shell just at the supercluster position. This site should be shifted along the direction, parallel to the galaxy minor axis (due to the general galaxy disk inclination), and also along the direction, parallel to the galaxy major axis (due to the suggested local plane inclination in relation to the NGC 6946 disc). The observed position of the break-through hole in the shell should be then somewhere between the major and minor axes of the complex image, which is indeed observed (Fig. 20).

The complicated evolution of the ISM under the influence of the stellar winds and supernovae of a supercluster was modeled in Tenorio-Tagle et al. (2006); they found that multiple shells within shells must be formed. Such a picture may correspond to our observed circumcluster and circumcomplex shells and also to structures, seen in the x, Vr plane in Figs. 19 and 21. Kinematical evidence for multiple shells around OB-associations and the related mechanism of triggered star formation was suggested by Chernin & Lozinskaya (2002). Of course the NGC 6946 supercluster is older than clusters considered in both models.

### 3.5.3 The analogy between the Hodge and the Gould complexes

Assuming that tilt and gas outflow exist in the NGC 6946 complex, we note that rather similar features were found in the Gould Belt. The angle of its plane to the Galactic one is 18º (Pöppel 1997; Perrot & Grenier, 2003; Elias et al. 2006), and two gas chimneys have been found to outflow in the opposite directions, orthogonal to the Gould Belt plane, to which the gas clouds and the younger stars concentrate (Lallement et al. 2003).

Thus we suggest that the Hodge complex might be a structure rather similar to the Gould Belt, a star complex which the Sun lies within. The latter was probably formed as a result of an impact of a supercloud (Wesselius & Feijes, 1973; Comeron & Torra, 1992, 1994; see also discussion in Pöppel, 1997) and this might also be the case for the Hodge complex. The important difference is, of course, the presence of the very rare object – a supercluster – in the NGC 6946 complex.

Is the Gould belt complex as a whole detached kinematically from the surrounding stuff of the Galaxy? Recently Bobylev (2006) has used 49 clusters to determined the mean motion of the Gould Belt complex relative to the local standard of rest, its velocity being 10.7 km/s and the velocity vector lying within the Galaxy plane. Such a direction of peculiar motion cannot be seen in NGC 6946. Seemingly nothing (apart from expansion) is known about the bulk motion of the ISM of the Gould complex. However, for the gas in the Hodge complex we may affirm that the systematic difference is small, if any; surely nothing is known about the velocities of its stars and clusters (apart from the SSC). The virial mass of the Gould belt (1.5 x $10^6$ suns, Bobylev 2006) is lower than the stellar mass of the Hodge complex (~$10^7$ suns, Larsen et al. 2002).

## 4 DISCUSSION

### 4.1 The hypotheses of impact

We consider first the suggestion that the morphology and gas velocity field of the Hodge complex is due to the impact of objects, coming either from the halo of the galaxy or from the intergalactic medium (IGM). As we have noted above, in some ways this complex is similar to the Gould belt complex, for which such an origin was suggested long ago.

There are more than a hundred HI holes in NGC 6946, some of which are accompanied by HI clouds with deviating velocities that are plausibly responsible for the respective hole formation (Boomsma et al. 2004; Fraternali et al. 2004; Boomsma 2007). This might be the case also for the hole #85 (Boomsma 2007), surrounding the Hodge complex and centered on the deep dip.

The shock ionization within the deep dip seems to agree with this suggestion, in spite of the fact that such ionization is expected to be observable for a rather short time. After the impact, long standing interactions of the shock waves, triggering star formation, may exist, as Santillán et al. (1999) have shown.

Some properties of the Hodge complex are really unique; therefore, the event which has triggered its formation might be also rare. Perhaps this is an example of a dark matter mini-halo impact which is able to trigger local star formation and produce the hole in the gas disc of a galaxy; the gravitational effects of “dark impact” are noted by the galaxy disc both before and after the impact itself (Bekki & Chiba, 2006). Low-mass dark matter haloes are suggested (Braun & Burton, 1999; Burton &Braun, 2000) to be associated with a population of compact high velocity HI clouds (CHVCs). Recently Thilker et al. (2004) concluded that over 80% of 95 HVC which they detected within 100 kpc around of M31 are of the CHVC class. Such mini-haloes/CHVCs might be quite numerous around large galaxies and associated with the smallest spheroidal dwarf galaxies (Kleyna et al. 2005).

The Hodge complex is located exactly in the centre of the "galactic coronal hole", the vast region, over which the magnetic field is oriented along the line of sight, i.e. at an inclination of ~60° to the plane of the galaxy (Beck, 1991, 2007). Might both the vast “coronal hole” and our complex be formed as a result of the passage of a dark mini-halo (coupled with baryonic matter) through the galaxy disk?

In the dark impact scenario, the impact would first provide the density enhancement necessary to trigger star formation and only later lead to the hole formation (Bekki & Chiba, 2006). The HI cloud velocities mostly follow the local HII velocities, as is seen in Fig. 19, and they might be the NGC 6946 clouds directed by the impacted dark matter’s gravitation. The shock ionization around the deep dip position points to the presence of baryonic matter in the impacted clouds, as is expected.

All HI velocity slices through the SSC position (Fig. 21) reveal HI clouds, the velocities of which deviate to lower values than the local galactic rotation. This is often observed near HI holes in NGC 6946 (Boomsma 2007), and is seen also in Fig. 16. We suggest that this might be due to cloud movements under gravitation of massive impacted bodies.

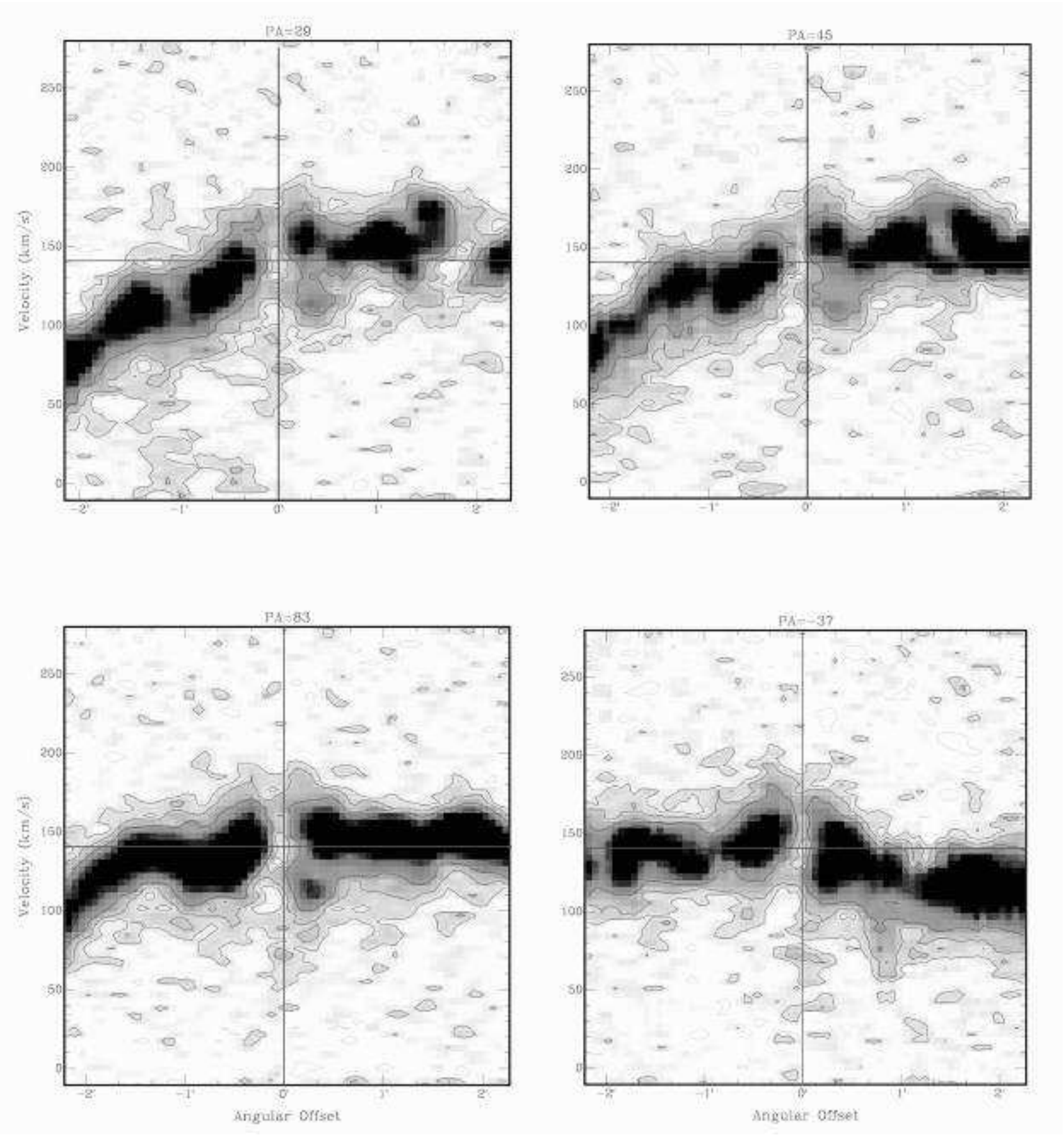

Figure 21. The HI velocity curves through the SSC position. The resolution is the same as shown in Fig. 14. Position angles are 29 (top left), 45 (top right), 83 (bottom left) and -37 (bottom right) degrees. The supercluster is in the center of the field-wide cross.

At any rate, in the giant expanding hole/supershell in M101, which was surely formed by stellar activity the cloud velocities deviate to both directions (Kamphuis et al. 1991). Recent studies suggest that subhaloes of dark matter inside large galaxies might be much more numerous than was believed ( Madau, 2007). The impact hypothesis is compatible with evidence for IGM accretion onto the NGC 6946 disk, such as the existence of a HI protuberance to the NW of NGC 6946, which might be the remnant of an accreted dwarf galaxy (Boomsma et al. 2004: Boomsma, 2007).

The impacting object, whether or not connected with dark matter, might be a compact dwarf galaxy, like one which was suggested to have triggered the formation of a star-forming complex in NGC 4559 (Soria et al. 2005). These authors noted that the $H_\alpha$ image suggests that the complex has a ring-like appearance and a possibly responsible satellite dwarf is seen nearby.

Recently, another galaxy, NGC 922, was suspected to form its very regular C-shaped eastern rim (quite similar to the western rim of the Hodge complex) as a result of a high-speed drop-through collision of a small galaxy, seen near NGC 922 at the side opposite to its eastern rim (Wong et al., 2006). The relative position and the arc-like rim's orientation there are the same as for our complex's western rim and the deep dip. The star-forming eastern rim of NGC 922 was suggested to have arisen as a result of a shock wave, expanded over NGC 922 after the collision. The same origin is possible for the western rim of the Hodge complex, and probably of the whole complex with its SSC and the regular patterns of young stars inside it.

The star forming ring within the peculiar galaxy Arp 10 was studied recently by Bizyaev, Moiseev, & Vorobyov (2007). They found clear evidence for the presence of an impacting small galaxy, projected still on Arp 10, and evidently responsible for the propagating star formation in the latter.

Anyway, we note that we do not see a potentially responsible galaxy near enough to NGC 6946.

### 4.2 Merging or projecting BCD galaxy

We have noted the striking similarity of the Hodge complex to the NGC 1705 BCD galaxy, both having an SSC and a size around 0.7 - 1 kpc. This was one of the reasons to suggest that the complex might be in fact a foreground or merging BCD galaxy, the other reason being the absence of a HI hole around the supercluster (Efremov et al. 2004).

Now, however, an HI hole has been found around the complex, though not centered on the SSC. Also, the similarity (outside the dip) between the velocities of the star complex (that of HI, HII and the SSC) and surrounding NGC 6946 gas, as well as the absence of evidence for any additional gas density in the complex, make this hypothesis rather improbable.

### 4.3 Nothing special?

The peculiar complex hosting the SSC might be just a result of processes common in spiral galaxies. A massive gas cloud is easier to form at the tip of a density wave spiral arm; later on it might undergo an instability event and form a first generation of stars (Elmegreen et al. 2000). The large amount of energy and momentum released into the ISM by several generations of stars (especially those in the SSC), might generate a hole/shell whose expansion is observed both in HI and ionized gas. A (one-sided) gas outflow might then be formed, as suggested in Section 3.5.2.

At any rate, the explanation of the deep dip feature in the velocity field is an issue not directly connected with the formation mechanism of the complex and the SSC. The source of energy for the rapid motion of HII and HI clouds inside the complex might well be the SSC, but this may tell nothing about the formation of the latter. It is necessary to explain why such complexes are not quite numerous at the tips of spiral arms. Lastly, the complex is at the tip of a short spur, not of a density wave spiral arm (see Fig. 1).

## 5 SUMMARY

Let us summarize the observed data on kinematics and flux distribution of ionized and neutral gas in the Hodge complex region. An oval depression in line-of-sight velocities of ionized gas, about 300 pc across, is seen to the east of the supercluster, the velocity of the ionized gas in its center being ~100 km/s blue-shifted. A symmetrical smaller red-shift of the line-of-sight velocity at both of the depression sides gives the HII velocity slice a crater-like appearance. The SSC is ~200 pc from the crater center and the latter (not the SSC) is near the center of a much larger HI hole.

The velocity field of HI gas reveals a gradient in the SE -- NW direction, which is compatible with gas expansion within the NGC 6946 plane, its center being again near the crater. In the intensity weighted HI velocity field there is no feature corresponding to the crater, because of the 13" resolution of HI data (equal to the crater size). However, HI clouds with large negative velocities are seen there in the velocity slices.

The peculiar complex hosts a supercluster and the velocity dip inside the encompassing expanding HI/HII hole. We cannot conclusively dterminewhich event was the first to trigger these features. The most simple explanation might be the origin of the SSC inside a high pressure complex (plausibly the result of an impact) and then the release of energy of its stars seen as the velocity dip rather far from the cluster, due to the inclined gas plane of the complex. The gas outflow might not be bi-lateral if the SSC is outside the mid-plane of the gas disc. The impact event might explain the regular arc-like shape of the complex's western rim and regular structures inside the complex.

In many respects – the SSC formation excluded – the scenario for the formation of the Hodge complex as a result of an impact event may resemble those suggested for the formation of the Gould Belt complex in the Milky Way galaxy. The tilt of the resulting complex plane after an oblique impact and the active star formation are the intrinsic properties of such a scenario (Comeron & Torra 1992, 1994). The formation as a result of a gas cloud impact has been suggested also for an isolated star complex in M83 (Comeron, 2001).

The SSC might be formed due to post-impact shock wave collisions (Chernin, Efremov & Voinovich 1995), which are suggested in Santillán et al. (1999), and connected high gas pressure, which is favorable for the formation of massive but still bound clusters (e.g. Elmegreen & Efremov 1997).

Both spots with maximal deviations of the ionized gas velocity are inside the HII hole at the complex's eastern part - they are the "crater" and a much smaller spot near the hole's NE edge. Both these spots are also those that demonstrate the maximal degree of shock excitation. In the framework of the impact hypothesis, the deviating gas velocities might be velocities of clouds which have passed through the galaxy disk. However, the shock excitation may also reveal the condition of gas clouds in the outflow from the surroundings of the supercluster. The crater-like velocity feature is then explained by the (one-sided) gas outflow, driven by pressure from the SSC, as was observed in the galaxy NGC 1705 (Meurer et al. 1992, 1998).

All in all, the impact origin seems to be the most plausible explanation for the Hodge complex, especially considering that it hosts a supercluster and a surprisingly regular C-shaped rim. Most of the detected features of its gas kinematics might be due to the stellar winds and SNe from the supercluster. We note in conclusion that until now the feature found 40 years ago in NGC 6946 is still the most enigmatic stellar system in the Local Universe, being similar in this relation only to the system of the giant star/cluster arcs in the LMC, also noted in the same paper by Hodge (1967).

Modeling of the impact of HI clouds and dark mini-haloes in terms of the observables (line-of-sight velocities versus coordinates on the sky) will be necessary to compare meaningfully the theoretical predictions with the observational data. We need also deeper and more wide-field stellar data as well as higher resolution data on HII, HI and CO fluxes and velocities. The issue of the origin of the features of the Hodge complex is crucial for understanding the contemporary formation of supermassive bound clusters.

**Acknowledgements**

This work is based on observational data obtained from the 6-m SAO RAS telescope financed by the Ministry of Science of Russia (registration # 01-43), and from the Westerbork Synthesis Radio Telescope which is operated by the Netherlands Foundation for Radio Astronomy with financial support from the Netherlands Foundation for the Advancement of Pure Research (N.W.O ), and also from the Gemini North telescope (AURA, USA).

YuNE is thankful to the supports from the Russian foundations RFBR (project 06-02-16077) and Sci. Schools (project 5290.2006.2). EJA would like to acknowledge the funding from MEC of Spain through grants AYA2004-05395, and AYA2004-08260-C03-02 (Estallidos), and from Consejería de Educacioʼn y Ciencia (Junta de Andalucía) through TIC-101.

We are deeply grateful to the referee for constructive comments and valuable advice. We acknowledge the useful discussions of some issues with K.Bekki and B.Elmegreen. Our thanks are due to G.Tracho, who assisted at Gemini North observations, and to M. Westmoquette for help in the analysis of its results.